\documentclass[fleqn,usenatbib]{mnras}

\usepackage{newtxtext,newtxmath}

\usepackage[T1]{fontenc}

\DeclareRobustCommand{\VAN}[3]{#2}
\let\VANthebibliography\thebibliography
\def\thebibliography{\DeclareRobustCommand{\VAN}[3]{##3}\VANthebibliography}

\usepackage{graphicx}	
\usepackage{amsmath}	
\usepackage{multirow}
\usepackage{subcaption}
\usepackage{comment}

\newcommand{\mbf}[1]{\mathbf{#1}}

\newcommand{\CBB}[1]{C^{BB,{#1}}_\ell}
\newcommand{\tCBB}[1]{\tilde{C}^{BB,{#1}}_\ell}

\newcommand{\pc}[3]{${#1}^{+{#2}}_{-{#3}}$}

\newcommand{\cc}[1]{\left({#1}\right)}
\newcommand{\rr}[1]{\left[{#1}\right]}
\newcommand{\be}{\begin{equation}}
\newcommand{\ee}{\end{equation}}
\def\bear#1\ear{\begin{align}#1\end{align}}

\newcommand{\f}{\frac}

\newcommand{\e}{\mathrm{e}}

\renewcommand{\mathbf}[1]{\mbox{\boldmath $#1$}}
\newcommand{\wj}[6]{\left(
                           \begin{array}{ccc}
        \! #1\! & #2\!  & #3\!\!  \\
        \! #4\! & #5\!  & #6\!\!
                           \end{array}
                   \right)}

\title[Disentangling patchy reionization from gravitational waves]{Disentangling patchy reionization signatures from primordial gravitational waves using CMB $E$-mode and $B$-mode polarization}

\author[Jain, Mukherjee \& Choudhury]{
Divesh Jain$^{1}$\thanks{djain@ncra.tifr.res.in}, 
Suvodip Mukherjee$^{2}$\thanks{suvodip@tifr.res.in},
and Tirthankar Roy Choudhury$^{1}$\thanks{tirth@ncra.tifr.res.in}
\\
$^{1}$ National Centre for Radio Astrophysics, Tata Institute of Fundamental Research, Pune 411007, India\\
$^{2}$ Department of Astronomy \& Astrophysics, Tata Institute of Fundamental Research, 1, Homi Bhabha Road, Colaba, Mumbai 400005, India\\
}
\date{\today}
\pubyear{2023}

\begin{document}
\label{firstpage}
\pagerange{\pageref{firstpage}--\pageref{lastpage}}
\maketitle

\begin{abstract}
The detection of large angular scale $B$-mode in the Cosmic Microwave Background (CMB) polarization signal will open a direct window into not only the primary CMB anisotropies caused by the primordial gravitational waves (PGW) originating in the epoch of inflation, but also the secondary anisotropies imprinted during the epoch of cosmic reionization. The existence of patchiness in the electron density during reionization produces a unique distortion in the CMB $B$-mode polarization, which can be  distinguished from the PGW signal with the aid of spatial frequency modes. In this work, we employ an $EB$ estimator by combining $E$-mode and $B$-mode polarization for the $\tau$ power spectrum signal generated in a photon-conserving semi-numerical reionization model called SCRIPT. We developed a Bayesian framework for the joint detection of the PGW and reionization signal from CMB observations and show the efficacy of this technique for upcoming CMB experiments. We find that, for our model, the $\tau$ power spectrum signal effectively tracks the inhomogeneous electron density field, allowing for robust constraints on the patchy $B$-mode signal. Further, our results indicate that employing the $EB$ estimator for the $\tau$ signal will facilitate ground-based CMB-S4 to detect the patchy $B$-mode signal at approximately $\geq 2\sigma$ confidence level while observations with space-based PICO will improve this detection to $\geq 3\sigma$ going as high as $\geq 7\sigma$ for extreme reionization models. These findings not only highlight the future potential of these experiments to provide an improved picture of the reionization process but also have important implications towards an unbiased measurement of $r$.

\end{abstract}

\begin{keywords}
cosmic background radiation -- dark ages -- reionization -- first stars -- cosmology: observations
\end{keywords}



\section{Introduction}

The epoch of reionization marks the period when the first stars and galaxies  ionized the cold and neutral hydrogen in the intergalactic medium to a warm and ionized one. It started around the redshift of $z\sim 20-30$ and ended at $z\sim 5-6$ \citep{2006AJ....132..117F,2015MNRAS.447.3402B,2018MNRAS.479.1055B,2019MNRAS.485L..24K,2021MNRAS.501.5782C}. Understanding reionization is important 
 for constructing a coherent narrative of cosmic evolution, as it provides insights into the nature of the first luminous sources, their clustering properties, and the intricate interplay between radiation and matter in the early universe \citep{2001PhR...349..125B,LoebFurlanetto+2013}. Observations of CMB anisotropies have begun to place important constraints on the details of the reionization process and indicate a reionization scenario that leads to a late and fast transition from a neutral to an ionized universe \citep{2020A&A...641A...6P}. Further, simulations, as well as analytical studies, suggest that the ionization fraction is spatially inhomogeneous leading to a patchy picture of reionization \citep{2000ApJ...530....1M,2001PhR...349..125B,2004ApJ...613....1F,2005MNRAS.363.1031F,2007ApJ...671....1T,2007ApJ...654...12Z,2011MNRAS.411..955M,2018MNRAS.481.3821C,2021MNRAS.500..232P}.  The inhomogeneous distribution of free electrons during this era impacts the temperature and polarization properties of the CMB by altering the line-of-sight optical depth statistics \citep{1998ApJ...508..435G,2000ApJ...529...12H}. The kinematic Sunyaev-Zeldovich (kSZ) signal arises from the Doppler shifting of photons scattering off moving ionized bubbles. Meanwhile, the large-scale ionization fluctuations lead to a net $B$-mode polarization signal as a result of inhomogeneous Thomson scattering of CMB photons. The strength of the kSZ signal \citep{Park_2013,2021MNRAS.500..232P,2022A&A...662A.122G,2022ApJ...927..186T,2023ApJ...943..138C} and $B$-mode signal \citep{2007ApJ...657....1M,2019MNRAS.486.2042M,2021MNRAS.500..232P,2021JCAP...01..003R} induced during reionization depends on the patchiness in the electron density and several efforts have been made to capture this through physics-driven simulation models. Recently, 
\cite{2021ApJ...908..199R} made the first statistically significant detection of kSZ. The fidelity of detection of these secondary CMB anisotropies has begun to place important constraints on both reionization history as well as the sources of ionizing photons \citep{2020MNRAS.499..550Q,2021MNRAS.501L...7C,2023ApJ...943..138C}.

Stage-4 CMB experiments will be targetting the detection of primordial $B$-mode polarization \citep{2018JLTP..193.1048S,2019arXiv190210541H,2019JCAP...02..056A,2019arXiv190704473A}, a characteristic imprint of primordial gravitational waves (PGW) on CMB, predicted by inflationary class of models. The amplitude for this signal is tied to a parameter $r$, called the tensor-to-scalar power spectrum ratio. An unbiased measurement of $r$ will potentially constrain the diverse landscape of inflationary theory \citep{2017JCAP...04..006S,2020A&A...641A..10P,2020EPJC...80.1163Q}. The latest constraint has been obtained by \cite{2022arXiv220316556B} at $r<0.035$ $(95\% \text{ C.L.})$. In addition to the detection of PGWs, high-fidelity large-scale-$B$ mode observations enable a unique window to detect the patchy $B$-mode polarization arising during reionization. Disentangling the patchy-$B$ mode from the primordial $B$-mode should be possible through optimal estimator construction at multipole range $\ell\sim 2-200$  \citep{2019MNRAS.486.2042M}. Within this range, the shape of patchy $B$-mode remains the same only to vary in amplitude depending on the choice of reionization scenario, while the shape of primordial signal is expected to be robust within the standard $\Lambda$-CDM cosmology \citep{2019MNRAS.486.2042M}. This would allow the possibility of inferring both primordial and patchy $B$-mode jointly from the CMB data. Even though the presence of secondary polarized radiation sources \citep{2016ARA&A..54..227K,2020A&A...641A...4P,2020A&A...641A..11P}, such as dust and synchrotron emission from our own galaxy and radio emission from extragalactic sources, makes observation of $B$-mode challenging, but, one of
the key aspects which make it possible to distinguish between CMB
and foregrounds is their distinguishable frequency spectrum between
a few tens of GHz to nearly THz frequency range \citep{2016ARA&A..54..227K}. 
However, the same is not true for patchy $B$-modes arising from reionization and we need to make sure that
the inferred value of tensor to scalar ratio is due to primordial gravitational waves and not due to patchy reionization. The patchy $B$-mode component of this polarization signal has been demonstrated to introduce a fractional bias in the measurement of the tensor-to-scalar power spectrum ratio, denoted as $r$ \citep{2019MNRAS.486.2042M,2021JCAP...01..003R,2023MNRAS.522.2901J} in context of sensitivities of the Stage-4 CMB experiments \citep{2018JLTP..193.1048S,2019arXiv190210541H,2019JCAP...02..056A,2019arXiv190704473A}.  This highlights the imperative of jointly constraining the patchy and primordial $B$-modes.

Patchy reionization contribution to $B$-mode signal can be traced either through its effect on the CMB power spectra or through the off-diagonal correlations induced by anisotropic or patchy $\tau(\hat{n})$. In this work, we focus on constraining the patchy $B$-mode through the combination of the above techniques, i.e., using CMB $B$-mode power spectrum in conjunction with reconstructing the patchy $\tau(\hat{n})$ estimate from off-diagonal correlations. Estimate for patchy $\tau(\hat{n})$ using minimum variance quadratic estimators was first proposed in \cite{Dvorkin:2008tf}. Employing a bubble-based prescription for reionization with characteristic bubble size of $5 \; {\rm Mpc}$ in WMAP cosmology \citep{2009ApJS..180..330K}, they estimated that the patchy reionization signal could be detected with $f^{-1/2}_{\rm sky}(S/N) \sim 14.7$ for an instrument specification with beam width $\Theta_{\rm f}=1$ arcmin and  polarization map depth $\Delta_P=0.5 \mu K$-arcmin. Revisiting the bubble-based  reionization prescription with updated star formation rate functions for high redshift galaxies, \cite{2018JCAP...05..014R} explored how the angular power spectrum of optical depth scales when changing parameters associated with bubble specification and their distribution. They concluded that for a CMB-S4 like instrument, a characteristic bubble radius of 5~Mpc would imply a $(S/N)\approx 4.8$ allowed by $\tau$ constraints from \cite{2016A&A...596A.108P}. Reconstruction of the patchy-$\tau$ has also been explored in the prospect of constraining the reionization timeline through three-point $\tau-$21~cm correlation statistics to forecast constraints on the width of reionization and Thomson scattering optical depth \citep{2013ApJ...779..124M}. However, it is important to note that simplistic spherical bubble-based prescriptions, employed in the above studies, become inaccurate when individual ionized bubbles begin to overlap, hence more realistic numerical methods need to be employed to explore the prospect of constraining patchy reionization using estimates of the patchy $\tau$ field. These realistic methods should capture the evolution of neutral hydrogen in the intergalactic medium (IGM) for large simulation boxes. Additionally, Bayesian inference studies require numerically efficient evaluation of reionization observables. In this regard, physically motivated semi-numerical models are a natural choice as they are numerically efficient in parameter space exploration while allowing us to track
relevant astrophysical parameters like Thomson scattering optical
depth $\tau$, free electron fraction, and more. Following our earlier works \citep{2019MNRAS.486.2042M,2021MNRAS.500..232P,2021MNRAS.501L...7C,2023MNRAS.522.2901J}, we simulate the patchy reionization using
an efficient explicitly photon-conserving semi-numerical model of
reionization, namely, Semi-numerical Code for ReIonization with
PhoTon-conservation (SCRIPT; \cite{2018MNRAS.481.3821C}). The advantage of this model is that it allows flexible parameterization of the ionizing sources and produces the
CMB signals at scales of our interest. In this study, we assess these within the self-consistent framework for CMB anisotropy evaluation as developed in \cite{2023MNRAS.522.2901J}.

In this study, we pursue three primary objectives. First, considering the need to develop optimal estimators for joint detection of patchy and primordial $B$-mode, we estimate the detectability of the optical depth power spectrum signal through off-diagonal correlation minimum variance estimators, as proposed in \citep{Dvorkin:2008tf}. We evaluate these estimators within the context of Stage-4 Cosmic Microwave Background (CMB) experiments, generated using a physically motivated semi-numerical model of reionization. Second,  we leverage Bayesian inference to forecast the insights this signal can offer regarding our physical model of reionization, as well as the extent to which it constrains the patchy $B$-mode signal as future CMB experiments come up. Third, we present the joint estimation of primordial and patchy $B$-mode capitalizing on the synergy between the $\tau$-power spectrum and $B$-mode power spectrum.

The paper is organized as follows: In Section \ref{sec:signalandsim} we describe patchy probes of reionization
and the simulation of reionization with which we evaluate this signal. We discuss the recovery of $\tau$-power spectrum signal with Stage-4 CMB experiments, namely CMB-S4 and PICO, in Section \ref{sec:taurecovery}. In Section \ref{sec:paramforecast}, we carry out parameter forecasts, in the context of these experiments, for our model of reionization using a combination of optical depth $\tau$ and the $\tau$-power spectrum. Finally, in Section \ref{sec:atau}, we exploit the synergy between $\tau$-power spectrum and scattering $B$-mode signal to forecast the detectability of the patchy reionization signal. Throughout the study, we have fixed the cosmological parameters to $[\Omega_m, \Omega_b, h, n_s, \sigma_8] = [0.308, 0.0482, 0.678, 0.961, 0.829]$ \citep{PlanckCollaboration2014} which is consistent with \cite{2020A&A...641A...6P}.

\section{Simulating Reionization signatures in the CMB}\label{sec:signalandsim}
During the reionization process, the patchy free electrons distribution at a comoving distance $\chi$ can be described as the sum of the mean free electron fraction and the fluctuations in the free electron fraction
\begin{equation}
    x_e\cc{\hat{n},\chi}=\bar{x}_e\cc{\chi}+\Delta x_e (\hat{n},\chi),
\end{equation}
where $x_e\cc{\hat{n},\chi} \equiv n_e\cc{\hat{n},\chi} / \bar{n}_H\cc{\chi}$ describes the ratio of the free electron comoving density to the mean comoving density of hydrogen and $\Delta x_e\cc{\hat{n},\chi} = \bar{x}_e\cc{\chi} \delta\cc{\hat{n},\chi}$, $\delta$ being the matter density contrast. Consequently, the optical depth to comoving distance $\chi$ along the line-of-sight $\hat{n}$ retains its direction dependence and can be expressed as:
\begin{equation}
    \tau(\hat{n},
    \chi)=\sigma_T \bar{n}_{H}  \int^\chi_0 \frac{d\chi^\prime}{{a^\prime}^2} \rr{\bar{x}_e(\chi^\prime)+\Delta x_e(\hat{n},\chi^\prime)},
\end{equation}
where $\sigma_T$ is the Thomson scattering cross-section. The optical depth to the last scattering $\tau(\hat{n})=\tau(\hat{n},\chi_{\rm LSS})$ surface can be evaluated by integrating the above integral till $\chi_{\rm LSS}$ corresponding to the redshift of last scattering surface given by $z_{\rm LSS}$.

The patchy distribution of free electrons impacts the statistics of the CMB dominantly via three principal mechanisms \citep{2000ApJ...529...12H}:
\begin{enumerate}
    \item Scattering (Temperature to $B$-mode): Thomson scattering of the local CMB temperature quadrupole off the patchy population of free electrons generates the large-scale $B$-mode polarization. Under Limber's approximation (valid at $\ell\gtrsim 30$),
    the patchy $B$-mode power is given as
\begin{eqnarray}
\begin{aligned}
\label{eq:BB_lim}
C_\ell^{BB,\mathrm{reion}} = \frac{6 \bar{n}^2_{H}\sigma^2_T}{100} \int  \f{d\chi}{a^4{\chi}^2}  &\e^{-2 \tau(\chi)} \times
\\&P_{ee}\left(k = \frac{l+1/2}{\chi} , \chi\right) \frac{Q_{\mathrm{RMS}}^2}{ {2}}.
\end{aligned}
\end{eqnarray}
Here, $P_{ee}(k)$ is the power spectrum of fluctuations in free electron fraction $x_e$. $Q_{\mathrm{RMS}}$ is the r.m.s of the primary quadrupole and is assumed to be constant at a value of $22~\mu K$ over the redshifts corresponding to the epoch of reionization \citep{Dvorkin:2008tf}.

    \item Screening ($E$-mode to $B$-mode): Patchy electron population modulates  both temperature and polarization anisotropies from the last scattering surface by a factor of $e^{-\tau(\hat{n})}$ as CMB photons traverse the reionization era. The direction-dependence of $\tau(\hat{n})$ introduces a phase factor, transforming the $E$-mode polarization arising from scalar density fluctuations into screened $B$-modes. Unlike the scattering effect, screening is dominant at small scales and can generate $B$-mode at these scales due to line-of-sight dependence on $\tau(\hat{n})$. Under flat-sky approximation \citep{PhysRevD.79.107302,2013PhRvD..87d7303G} the screening $B$-mode power spectrum is given as 
\bear
C_\ell^{BB,\mathrm{scr}}=\int \frac{d^2\ell^\prime}{{2\pi}^2}C^{EE}_{\ell^\prime}C^{\tau\tau}_{|\ell-\ell^\prime|}\sin 2\phi_{\ell^\prime}
\ear

    \item kSZ (Ionized momentum anisotropy to Temperature): Doppler shifting of CMB photons as they scatter off ionized regions with net bulk velocity introduces a new temperature anisotropy called the patchy kinetic-Sunyaev Zeldovich signal. Under Limber approximation, the angular power spectrum of kSZ signal sourced from the patchy reionization era is given as \citep{PhysRevLett.88.211301,Park_2013,2021MNRAS.500..232P}. \\

\begin{eqnarray}
\begin{aligned}
C_\ell^{\mathrm{kSZ,reion}} = \left(\sigma_T \bar{n}_{H} T_0\right)^2 \int d\chi \f{\e^{-2 \tau(\chi)}}{a^4\chi^2}~\f{P_{q_\perp}(k = l+1/2/\chi,\chi)}{2}.
\label{eq:ksz_lim}
\end{aligned}
\end{eqnarray}
Here, $P_{q_\perp}$ is the power spectrum of transverse component of the Fourier transform of the momentum field $\mbf{q}(\mbf{k},z)$.  In this work, we do not consider the prospect of using screening of CMB anisotropies and the kSZ signal to constrain our model of reionization, but, we provide the formulation for completeness on the discussion of reionization imprints on CMB.
\end{enumerate}

\subsection{The optical depth power spectrum signal}

 The direction dependence of $\tau(\hat{n})= \sum_{LM} a^\tau_{LM}Y_{LM}(\hat n)$ allows it to be decomposed in spherical harmonics coefficients $a^\tau_{L M}$, and the corresponding angular power spectrum  under Limber's approximation can then be given as  \citep{Dvorkin:2008tf}

\be 
C^{\tau\tau}_{L}=\cc{\sigma_T \bar{n}_{H,0}}^2 \int \frac{d\chi }{a^4 \chi^2}P_{ee}\left(k=\frac{L+1/2}{\chi},\chi\right),
\ee 
Under flat-sky approximation, \cite{PhysRevD.79.107302} provide a scaling relation between the above $\tau$-power spectrum and  
the scattering $B$-mode signal and is given by 
\begin{equation}\label{eq:tautauBB}
    C^{BB,{\rm reion}}_{L} \approx \frac{3}{100}C^{\tau\tau}_{L}Q^2_{\rm rms}e^{-2\tau}
\end{equation}
In this work, we exploit this relationship when we discuss the extraction of the patchy $B$-mode signal from the observed $B$-mode power.

\subsubsection{Noise modelling}
The patchy $\tau(\hat{n},\chi)$ modifies both polarization and temperature properties of the CMB photon population. Therefore, in principle, all six temperature and polarization correlations can be used to reconstruct the estimator for the field $\tau(\hat{n})$. But, the reconstruction using the EB minimum variance quadratic estimator will yield the best sensitivity to observe this signal \citep{Dvorkin:2008tf}, as it is not contaminated by the primary CMB anisotropy in absence of any parity violating effects. Therefore, in our analysis, we employ the optical depth reconstruction noise, $N^{\tau\tau}_{L}$ corresponding only for the $EB$ estimator, given by
\begin{equation}
    N^{\tau\tau}_{L}=\left[\frac{1}{(2L+1)} \sum^{\ell_{\rm max}}_{\ell_1\ell_2} \frac{|\Gamma_{\ell_1\ell_2L}^{EB(\tau)}|^2}{\left( C^{EE}_{\ell_1}+N^{EE}_{\ell_1}\right)\left( C^{BB}_{\ell_2}+N^{BB}_{\ell_2}\right)} \right]^{-1}
\end{equation}
The $C^{BB}_{\ell}$ and $C^{EE}_{\ell}$ are the total CMB $E$ \& $B$-mode power spectrum and $N^{BB}_{\ell}$ \& $N^{EE}_{\ell}$ are the instrument noise spectrum for $E$ \& $B$-mode polarization. The instrumental power spectrum which can be written as
\begin{equation}
    N^{XX}_{\ell}=\Delta^2_P \exp \rr{\frac{\ell(\ell+1)\Theta^2_f}{8 \ln{2}}},
\end{equation}
where $XX \in \{EE, BB\}$. $\Delta_p$
is the noise of the detector for polarization in $\mu$K-arcmin and holds the relation  $\Delta_p=\sqrt{2}\Delta_T$ , $\Delta_T$ being the detector noise for temperature observations. $\Theta_f$ is the full width half maxima (FWHM) of the beam in arcmin units.

\begin{figure}
    \centering
    \includegraphics[width=\columnwidth]{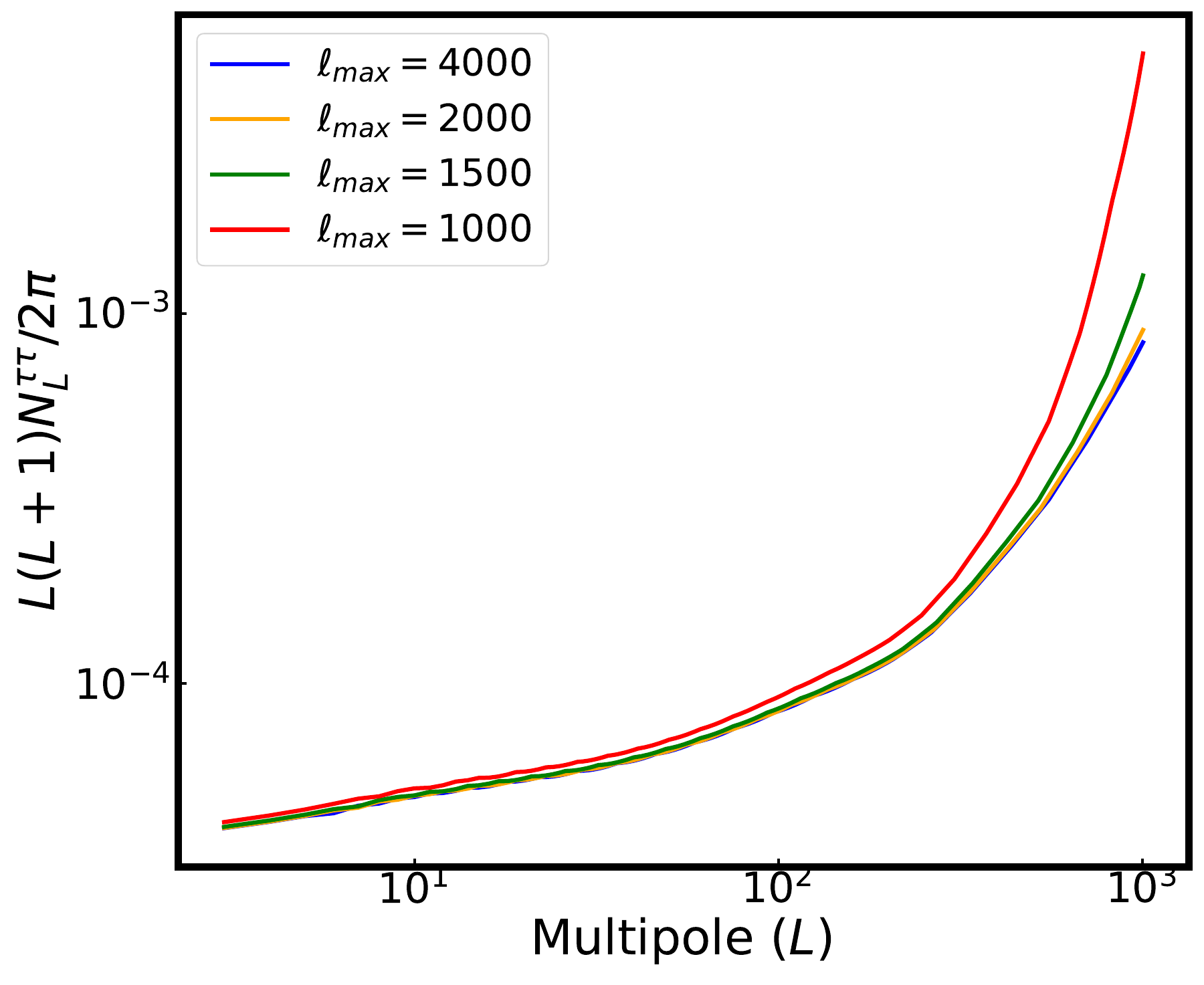}
    \caption{The optical depth reconstruction noise spectra $N^{\tau\tau}_{\ell}$ has been plotted for different $\ell_{max}$ for an experiment with $\Delta_p=\sqrt{2} \mu K$-arcmin and beam width of $\Theta_f=1 {\rm arcmin}$ }
    \label{fig:nelllimit}
\end{figure}

The EB coupling $\Gamma^{EB(\tau)}_{\ell_1\ell_2L}$ (valid for $L>10$) can be written as
\begin{eqnarray}
\begin{aligned}
        \Gamma^{EB(\tau)}_{\ell_1\ell_2 L}=-\frac{C^{E_0E_0}_{\ell_1}}{2i}\sqrt{\frac{(2\ell_1+1)(2\ell_2+1)(2L+1)}{4\pi}}\\\times \left[ \wj{\ell_1}{\ell_2}{L}{-2}{+2}{0} -\wj{\ell_1}{\ell_2}{L}{+2}{-2}{0}  \right],
\end{aligned}
\end{eqnarray}
where $C^{E_0E_0}_\ell$ is the $E$-mode power spectrum without the patchy reionization contribution.
The second line in the above equation requires the evaluation of Wigner-3j matrices. In order to determine the $\ell_{\rm max}$  for reconstruction noise $N^{\tau\tau}_{L}$, we plot $N^{\tau\tau}_{L}$ for various values of $\ell_{\rm max}$ in Figure \ref{fig:nelllimit}. We find that for $\ell_{\rm max}=2000$ convergence in the optical depth reconstruction noise spectrum is obtained with $\ell_{\rm max}=4000$. For further studies, we limit our evaluation to $\ell_{\rm max}=2000$. The analysis method developed here can be trivially extended to higher values of $L$.

\subsection{Simulating patchy effects of reionization using SCRIPT}

In the context of Bayesian inference, an efficient evaluation of observables is essential for each sampled set of free parameters. A physically motivated semi-numerical model of reionization is apt for this purpose as it helps us to track astrophysical observables while being numerically efficient. Hence, for this work, we use a semi-numerical scheme of reionization called \texttt{SCRIPT} for the efficient generation of ionization maps.

Semi-numerical Code for ReIonization with PhoTon-conservation, abbreviated as \texttt{SCRIPT}, is a semi-numerical scheme of reionization that is explicitly photon-conserving \citep{2018MNRAS.481.3821C}. Other than its time efficiency, \texttt{SCRIPT} generates power spectrums that are convergent at large scales across map resolutions. This feature differentiates \texttt{SCRIPT} from other semi-numerical models of reionization based on the popular excursion set approach \citep{2007ApJ...669..663M,2011MNRAS.411..955M,2008MNRAS.386.1683G}, preventing any inference bias when opting to work with coarser map resolutions.

To generate patchy CMB signals, we need ionization maps at redshifts of reionization. The first step is to generate dark matter snapshots at these redshifts. For a fixed set of cosmological parameters, we generate dark matter snapshots at $\Delta z=0.1$ for redshifts $5 \leq z \leq 20$ employing the 2LPT prescription in MUSIC \citep{hahn2011multi} for  box length of $512 ~ h^{-1}$ Mpc with $512^3$ particles. The collapsed mass fraction in haloes is computed using a subgrid prescription based on the conditional ellipsoidal collapse model \citep{2002MNRAS.329...61S}, see \cite{2018MNRAS.481.3821C} for more details of the method.

 In order to generate an ionization map at a redshift, \texttt{SCRIPT} requires two input parameters $M_{\rm min}$, the minimum mass of haloes that can host ionizing sources and $\zeta$, the effective ionizing efficiency of these sources. Using these parameters \texttt{SCRIPT} generates a map of ionized hydrogen fraction $x_{\mathrm{HII}}(\mbf{x},z)$. For this study, our parameter of interest is the free electron fraction
\be
x_e(\mbf{x},z) = \chi_{\mathrm{He}}~x_{\mathrm{HII}}(\mbf{x},z)~\Delta(\mbf{x},z),
\ee
where, $\chi_{\mathrm{He}}$ is the correction factor to account for free electrons from ionized Helium and $\Delta(\mbf{x},z)$ corresponds to the dark matter overdensity. In our analysis, we consider $\chi_{\mathrm{He}}=1.08$ for $z>3$ corresponding to contribution from singly-ionized Helium and $\chi_{\mathrm{He}}=1.16$ for $z<3$ to account for free electron contribution from doubly ionized Helium. To enable us to capture the small-scale inhomogeneities, ionization maps using \texttt{SCRIPT} are generated with the best possible resolution of $2 ~ h^{-1}$ Mpc. 

Modelling the reionization and hence the emerging CMB anisotropies, is contingent on the parameterization we assume for $M_{\rm min}$ and $\zeta$ across redshift. As it is not exactly clear how ionizing sources evolve in the reionization era, we assume an intuitive redshift-based power-law model for $M_{\rm min}$ and $\zeta$. The parameterization is thus considered as following
\be
\zeta(z) = \zeta_0 \left(\f{1 + z}{9}\right)^{\alpha_{\zeta}},~~ M_{\mathrm{min}}(z) = M_{\mathrm{min},0} \left(\f{1 + z}{9}\right)^{\alpha_{M}},
\ee
Here, $M_{\mathrm{min},0}$ is the minimum mass of haloes that can contribute to the ionizing process at redshift $z=8$ while $\zeta_0$ is the ionizing efficiency of these sources at $z=8$. The parameters $\alpha_M$ and $\alpha_\zeta$ correspond to indices of the power law. Therefore, the reionization process can be completely described by the four free parameters $\mathbf{\theta}\equiv[\log (\zeta_0),\log M_{\mathrm{min},0},\alpha_\zeta,\alpha_M]$. Given these four parameters, one can compute ionization maps with \texttt{SCRIPT}. Finally, the evaluation of the patchy reionization signals considered in this work requires an evaluation of the power spectrum of electron density fluctuations from the ionization maps. The scattering effect is dominant at large scales, the wave modes corresponding to which can be smaller than the smallest wave modes available in the simulation box, we direct the readers to \cite{2019MNRAS.486.2042M,2023MNRAS.522.2901J} for details on the evaluation of electron density fluctuation power spectrum at wavemodes smaller than $k_{\rm box}=2\pi/{L_{\rm box}}$. 

During noise modelling to detect the $\tau$ power spectrum in Section \ref{sec:taurecovery} and \ref{sec:paramforecast} and Bayesian inference to recover the amplitude of patchy reionization signal in Section \ref{sec:atau}, we  would need to evaluate the total $B$-mode power spectrum. The total $B$-mode contribution comes from the combination of primordial and lensing $B$-mode along with patchy $B$-mode from reionization. These signals are generally evaluated using an efficient Boltzmann solver for CMB anisotropies like \texttt{CAMB} \citep{2000ApJ...538..473L,2012JCAP...04..027H}. However, in its default setup \texttt{CAMB} assumes a $\tanh$ reionization scenario, which is rather a restrictive choice of reionization modelling. To ensure consistency in modelling both primordial and patchy reionization signals, we modified \texttt{CAMB} to accommodate varied reionization histories \citep{2023MNRAS.522.2901J}. In particular, we modify the \texttt{reionization.f90} code to take ionization history computed with \texttt{SCRIPT} as an input. This allows for the evaluation of each component of CMB anisotropies with a consistent reionization scenario predicted by our physical model.

\begin{figure}
	\includegraphics[width=\columnwidth]{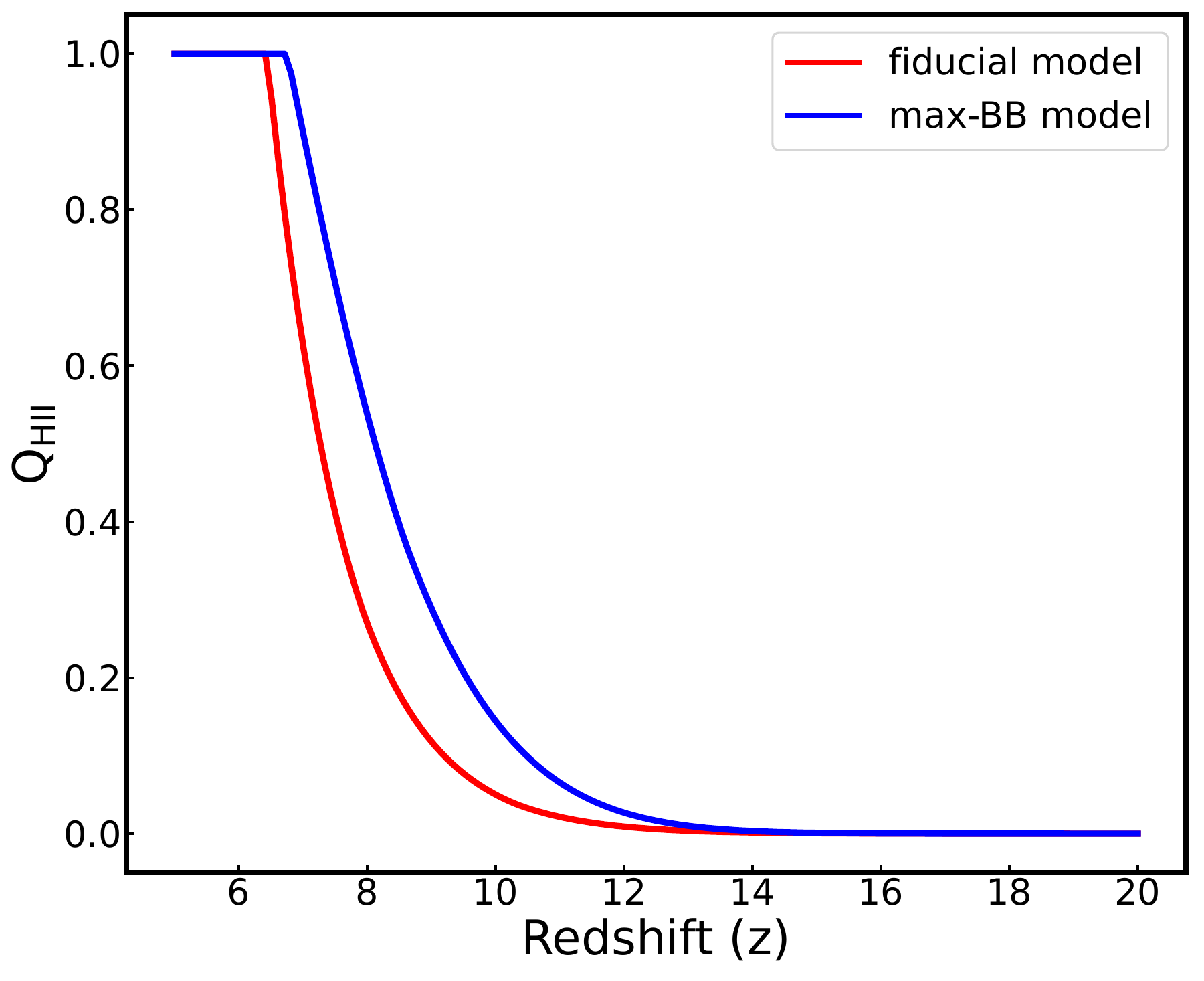}
    \caption{Redshift evolution of mass-averaged ionized fraction $Q_{\rm{HII}}(z)$ for the fiducial and the max-BB model of reionization considered from the work \citep{2023MNRAS.522.2901J}}
    \label{fig:ionizationfrac}
\end{figure}

Finally, for forecasting studies in this work, we require a model of reionization, that can best capture our understanding of reionization. Thus, we consider the best-fit model of reionization as the fiducial model of reionization, obtained corresponding to the Bayesian inference carried out on the above model of reionization in \cite{2023MNRAS.522.2901J} using recent CMB measurements. The inference used constraints on  $\tau=0.054$ with $\sigma^{\mathrm{obs}}_\tau=0.007$ \citep{2020A&A...641A...6P} and the kSZ signal  \citep{2021ApJ...908..199R} at $D^{\mathrm{kSZ,obs}}_{\ell=3000}\equiv\ell(\ell+1)C^{\mathrm{kSZ,obs}}_{\ell=3000} =3 ~\mu K^2$ with a $\sigma^{\mathrm{kSZ,obs}}_{\ell=3000}=1~\mu K^2$. The best-fit model parameterization was obtained to be $\rr{\log (M_{\mathrm{min,0}})=9.73, \log (\zeta_0)=1.58, \alpha_M=-2.06, \alpha_\zeta=-2.01}$. The value of $\tau$ for this fiducial model is $0.0540$ and the $B$-mode power at multipole of 200 is $D^{BB}_{\ell=200}=7.03 \; {\rm nK^2}$. The redshift evolution of the global mass-averaged ionization fraction $Q_{\rm{HII}}(z) \equiv \langle x_{\mathrm{HII}}(\mbf{x},z)~\Delta(\mbf{x},z) \rangle$ is shown in Figure \ref{fig:ionizationfrac} in red curve. In Section \ref{sec:atau}, when we discuss the extraction of the patchy $B$-mode signal, we also consider one extreme model of reionization that corresponds to the highest $B$-mode power allowed by the current CMB estimates \citep{2023MNRAS.522.2901J} called the max-$BB$ model of reionization. The source parameterization of the model is  $\rr{\log (M_{\mathrm{min,0}})=10.39, \log (\zeta_0)=2.48, \alpha_M=-0.76, \alpha_\zeta=3.58}$, the redshift evolution of ionization fraction is presented by the blue curve in Figure \ref{fig:ionizationfrac}. This model corresponds to reionization with an optical depth $\tau=0.0627$ and $B$-mode power $D^{BB}_{\ell=200}=18.41 \; {\rm nK^2}$.  

\begin{figure}
    \centering
    \includegraphics[width=\columnwidth]{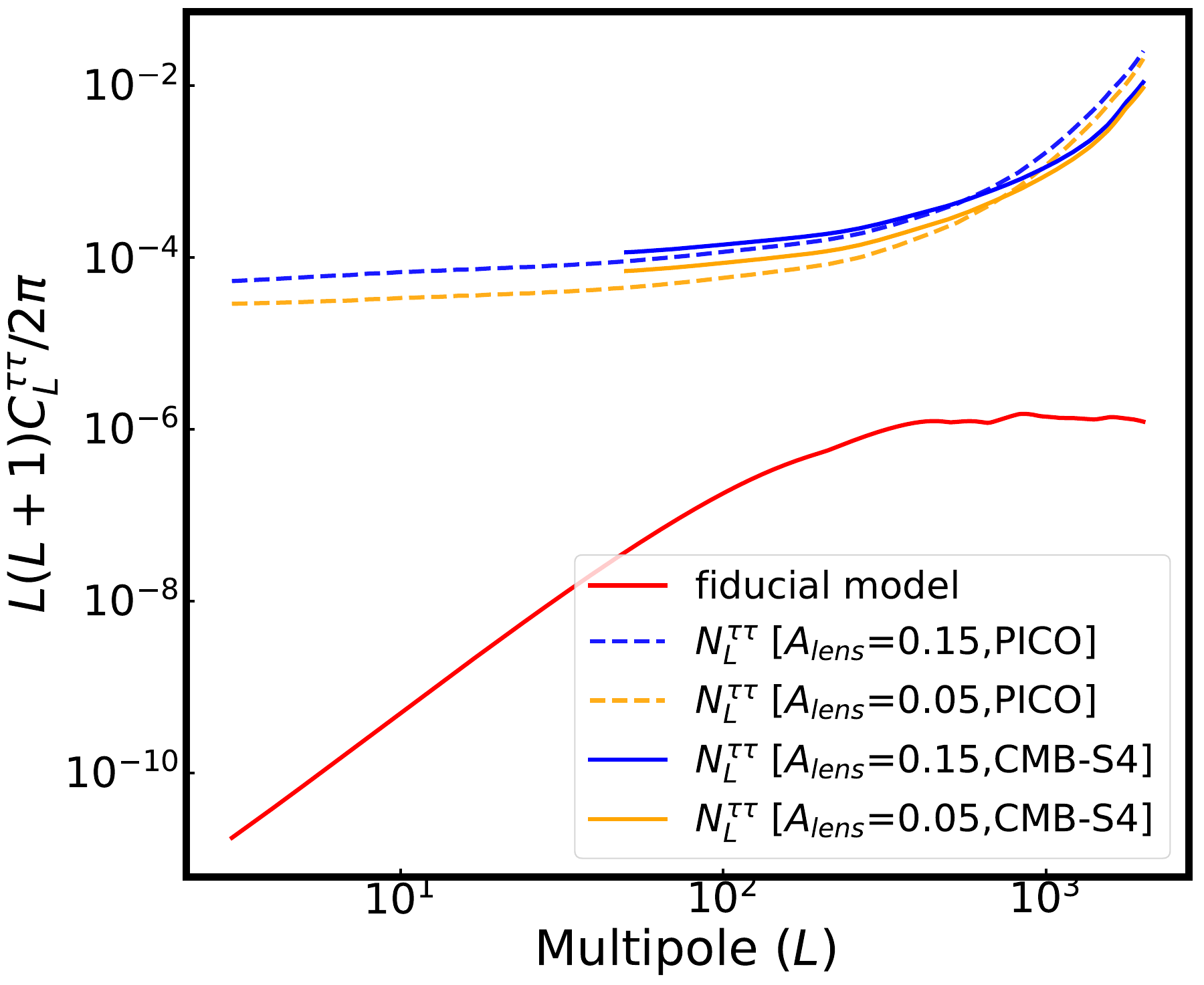}
    \caption{$\tau$-power spectrum for the fiducial model of reionization has been presented in solid red curve. The noise power spectrum at different delensing scenarios corresponding to observations with CMB-S4 and PICO has been presented in blue and yellow curves. }
    \label{fig:snbestfit}
\end{figure}

\begin{figure}
    \centering
    \includegraphics[width=\columnwidth]{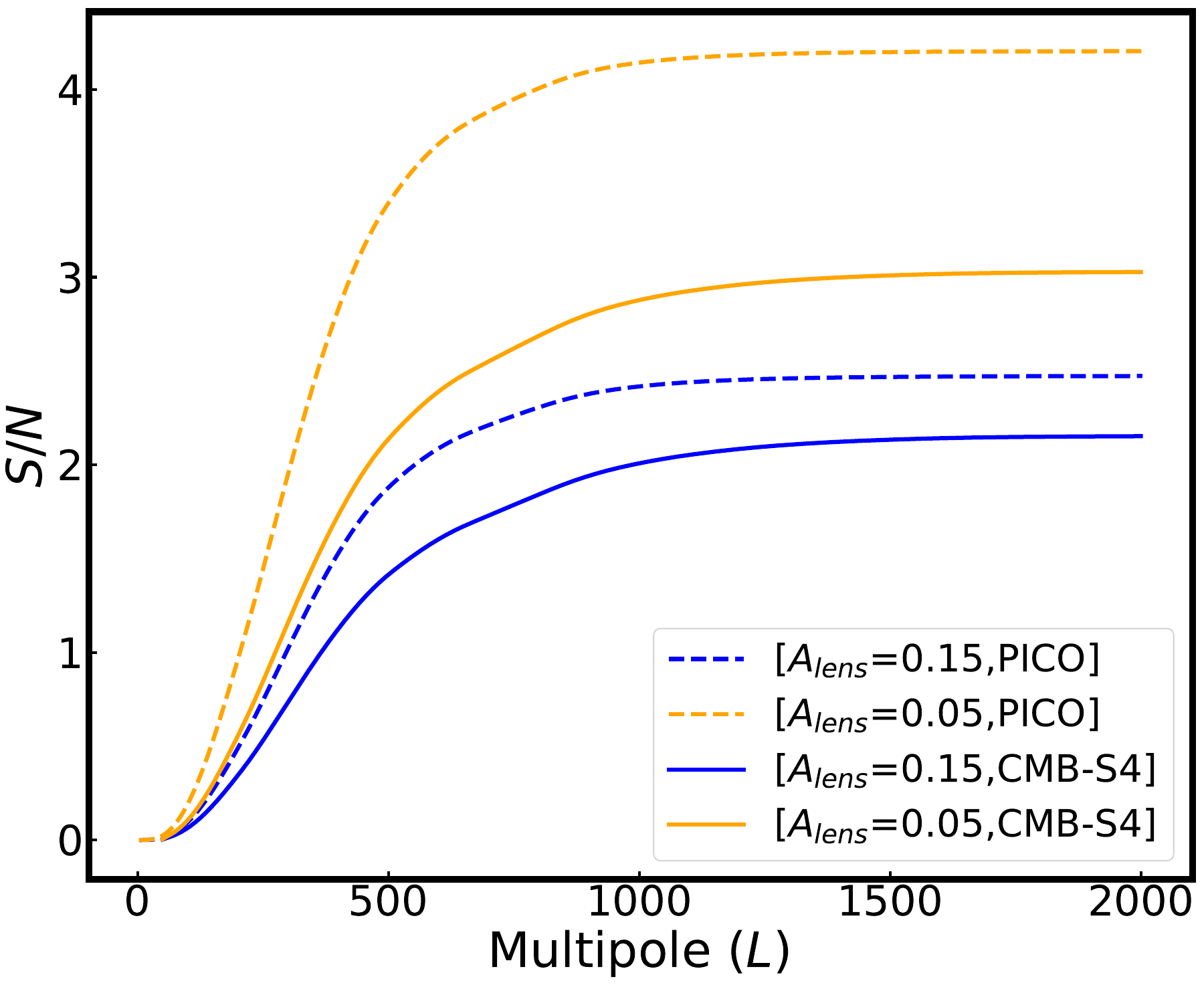}
    \caption{The cumulative signal-to-noise ratio corresponding to the fiducial model for different delensing criteria corresponding to polarization observation with CMB-S4  and PICO}
    \label{fig:sbynbestfit}
\end{figure}

\section{recovering optical depth angular power spectrum with CMB-S4 and PICO}\label{sec:taurecovery}
The most sensitive $\tau$-reconstruction is anticipated to be achieved using polarization maps (employing $EB$ estimator) from upcoming CMB experiments such as Simons Observatory \citep{SimonsObservatory:2018koc}, CMB-S4 \citep{2019arXiv190704473A}, LiteBIRD \citep{LiteBIRD:2020khw}, PICO \citep{2019arXiv190210541H}, CMB-HD\citep{CMB-HD:2022bsz}. In this analysis we focus on LiteBird \citep{LiteBIRD:2020khw}, CMB-S4 \citep{2019arXiv190704473A} and PICO \citep{2019arXiv190210541H}. For observations with CMB-S4 we consider  $\Delta_P=\sqrt{2}~\mu$K-arcmin with $\Theta_f=1.0$ arcmin and $f_{\rm sky}=0.7$ while for observations with PICO we consider polarization map depth of $\Delta_P=0.87~\mu$K-arcmin with $\Theta_f=7.9$ arcmin and $f_{\rm sky}=1.0$.

For our fiducial reionization model, the $\tau$-power spectrum signal along with the reconstruction noise for various delensing scenarios corresponding to CMB-S4 and PICO are shown in Figure \ref{fig:snbestfit}. While computing $N^{\tau\tau}_{L}$, the primordial $B$-mode power spectrum is set corresponding to $r=5\times 10^{-4}$ and galactic foreground contribution is neglected. The signal-to-noise ratio for the $\tau$-power spectrum signal can then be given by:
\begin{equation}
    {\rm (S/N)}^2=\frac{f_{\rm sky}}{2}\sum^{L_{max}}_{L_{min}}  (2L+1)\cc{\frac{C^{\tau\tau}_{L}}{N^{\tau\tau}_{L}}}^2,
\end{equation}

\begin{figure}
    \centering
    \includegraphics[width=\columnwidth]{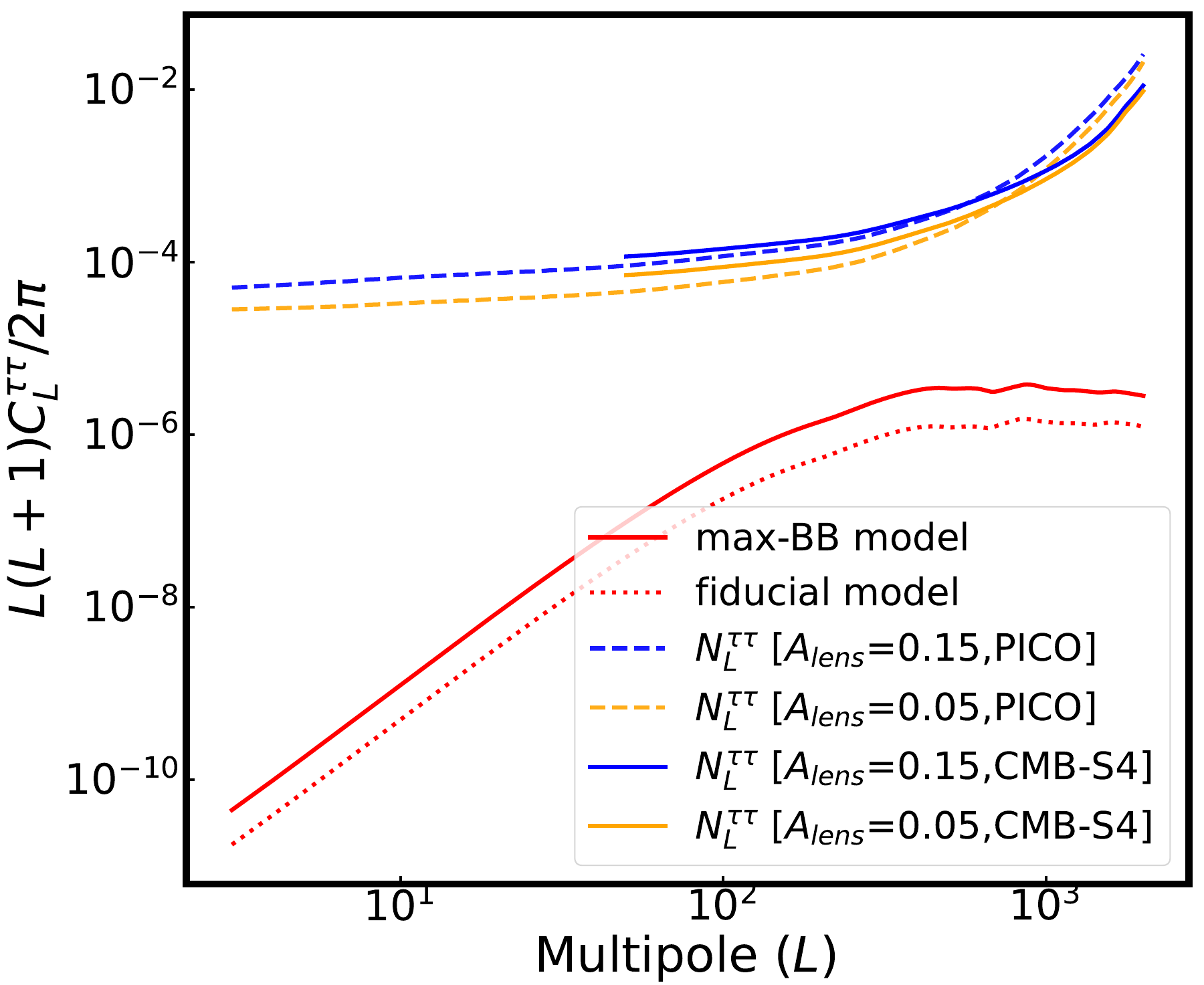}
    \caption{$\tau$-power spectrum for the max-BB model of reionization has been presented in solid red curve. For reference, the $\tau$-power spectrum for the fiducial model has been presented in a dotted red curve. Noise power at different delensing scenarios corresponding to the max-BB model has been presented corresponding to observations with CMB-S4 and PICO in blue and yellow curves.}
    \label{fig:signal3sigma}
\end{figure}

\begin{figure}
    \centering
    \includegraphics[width=\columnwidth]{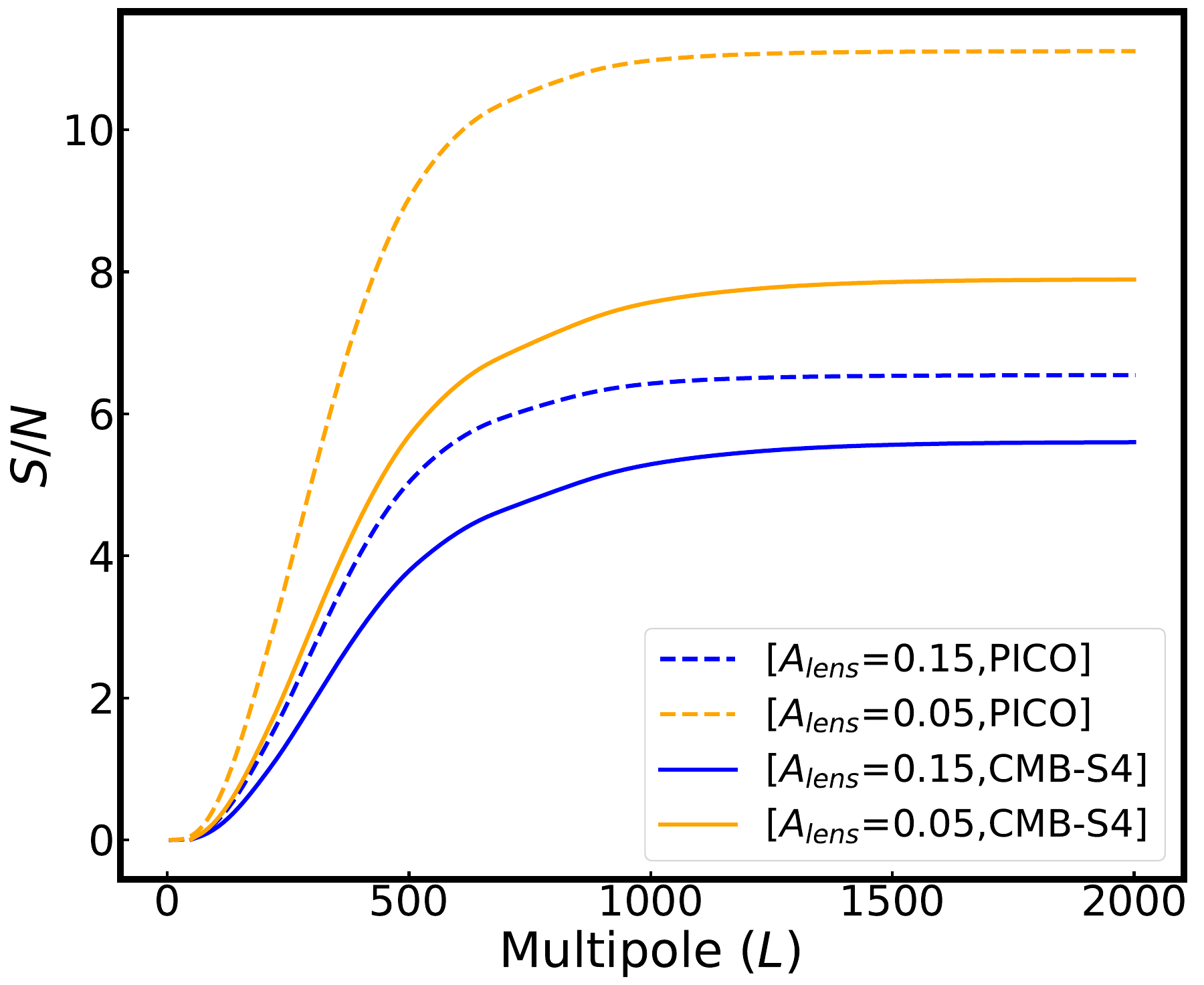}
    \caption{The cumulative signal-to-noise ratio corresponding to the max-BB model for different delensing criteria corresponding to polarization observation with CMB-S4 and PICO.}
    \label{fig:sn3sigma}
\end{figure}

As we will be considering the estimation of power in patchy-$\tau$ fields using scattering $B$-mode contribution we restrict our analysis to $L_{\rm max}\sim 2000$. For CMB-S4, we consider $L_{\rm min}=50$ and $L_{\rm max}=2050$, while for PICO, we consider $L_{\rm min}=2$ and $L_{\rm max}=2002$. The cumulative signal-to-noise corresponding to CMB-S4 and PICO has been shown in Figure \ref{fig:sbynbestfit}. From the cumulative S/N plot, we can infer that a $\gtrsim 3\sigma$ detection for the $\tau$-power spectrum is possible for both the instruments at a $95\%$ delensing scenario. Hence, for further analysis, we restrict ourselves to a $95\%$ delensing scenario when considering the fiducial reionization model. In Figure \ref{fig:signal3sigma}, we present the $\tau$-power spectrum for the max-BB model of reionization, the cumulative S/N for which has been shown in Figure \ref{fig:sn3sigma}. As the cumulative S/N for max-BB model is $>3$ for both $85\%$ and $95\%$ delensing scenarios, we consider both delensing scenarios when discussing the detectability of patchy $B$-mode signal for max-BB reionization model in Section \ref{sec:atau}.

\begin{table}
    \def\arraystretch{1.6}
    \centering
    \caption{Specifications of priors for the free parameters used during Bayesian inference of reionization parameters}
    \begin{tabular}{|c|c|c|}
    \hline
    Parameter & range & nature \\
    \hline
    $\log (\zeta_{\rm 0})$     & [0, $\infty$] & uniform\\
    $\log (M_{\rm min,0})$  & [7.0, 11.0] & uniform \\
    $\alpha_\zeta$ & [-$\infty$, $\infty$] & uniform\\
    $\alpha_M$ & [-$\infty$, 0] & uniform\\
    \hline
    \end{tabular}
    \label{tab:paramprior}
\end{table}

\begin{table}
    \centering
    \def\arraystretch{1.5}
    \caption{Parameter forecasts ($68\%$ limits) on free and derived parameters of the reionization model have been presented corresponding to different combinations of data sets }
    \begin{tabular}{|c|c|c|c|c|c|c|}
    \hline

    Parameters  & Input    & LB & LB+S4$\tau\tau$ & LB+PICO$\tau\tau$  \\
    \hline
    $\log M_{\rm min,0}$ & 9.73 & \pc{9.66}{0.99}{0.49}&\pc{9.70}{0.96}{0.49} &\pc{9.70}{0.96}{0.46}\\
    $\log \zeta_0$& 1.58 &\pc{1.63}{0.46}{0.64}&\pc{1.66}{0.44}{0.64}&  \pc{1.66}{0.46}{0.61}\\
    $\alpha_\zeta$ & -2.01 &\pc{-3.42}{2.37}{2.80}&\pc{-3.86}{2.44}{2.35}&\pc{-3.87}{2.41}{2.34}\\
    $\alpha_M$&   -2.06 &$>-2.73$&$>-2.87$&$>-2.87$\\
    \hline
    $z_{25}$ & 8.09 &\pc{8.13}{0.22}{0.29}&\pc{8.10}{0.22}{0.24}& \pc{8.11}{0.21}{0.26}\\
    $z_{50}$ & 7.27 &\pc{7.34}{0.27}{0.25}&\pc{7.35}{0.28}{0.23}& \pc{7.36}{0.23}{0.24}\\
    $z_{75}$ & 6.78 &\pc{6.87}{0.47}{0.31}&\pc{6.92}{0.45}{0.27}&\pc{6.93}{0.44}{0.30}\\ 
    $\Delta_z$ & 1.31 &\pc{1.25}{0.25}{0.62}&\pc{1.17}{0.23}{0.54}&\pc{1.18}{0.20}{0.51}\\ 
    $\tau$ & 0.054 &\pc{0.0545}{0.0019}{0.0020}&\pc{0.0545}{0.0019}{0.0020}&\pc{0.0546}{0.0019}{0.0020}\\ 
    $D^{BB}_{\ell=200}  {\rm (nK^2)}$ & 7.03 &\pc{6.71}{0.85}{2.93}&\pc{6.27}{0.98}{2.11}&\pc{6.25}{1.10}{1.81}\\ 
    $D^{\tau\tau}_{L=400}\times {\rm 10^{6}}$ & 1.19 &\pc{1.14}{0.15}{1.50}&\pc{1.07}{0.17}{0.36}&\pc{1.08}{0.19}{0.32}\\
    \hline
    \end{tabular}
    \label{tab:fullparamconst_bestfit_move}
\end{table}
\section{Parameter Forecasts}\label{sec:paramforecast}
In this section, we forecast constraints on reionization parameters using a combination of CMB probes of the mean value of optical depth
$\tau$ and $\tau$-power spectrum $C^{\tau\tau}_L$. This approach will provide us with the means to determine the best possible constraints on ionizing source properties and reionization histories which will be enabled by the Stage-4 CMB polarization data sets.

Stage-4 CMB experiments will make high-fidelity polarization observations, enabling extremely sensitive measurements of CMB probes of reionization. The upcoming experiments like LiteBIRD and PICO aim to make $E$-mode polarization observations at the reionization bump ($\ell<10$),  to constraint $\tau$ at $\sigma_\tau=0.002$. In this work, when projecting forecasts with LiteBIRD we consider $\sigma^{ obs}_\tau=0.002$. For, $\tau$ power spectrum data-sets based constraints we consider the instrument specification discussed in Section \ref{sec:taurecovery}.  Considering the upcoming mission timelines, we propose the following combination of mock data sets to forecast constraints on the reionization model:

\begin{figure*}
    \centering
    \includegraphics[width=\linewidth]{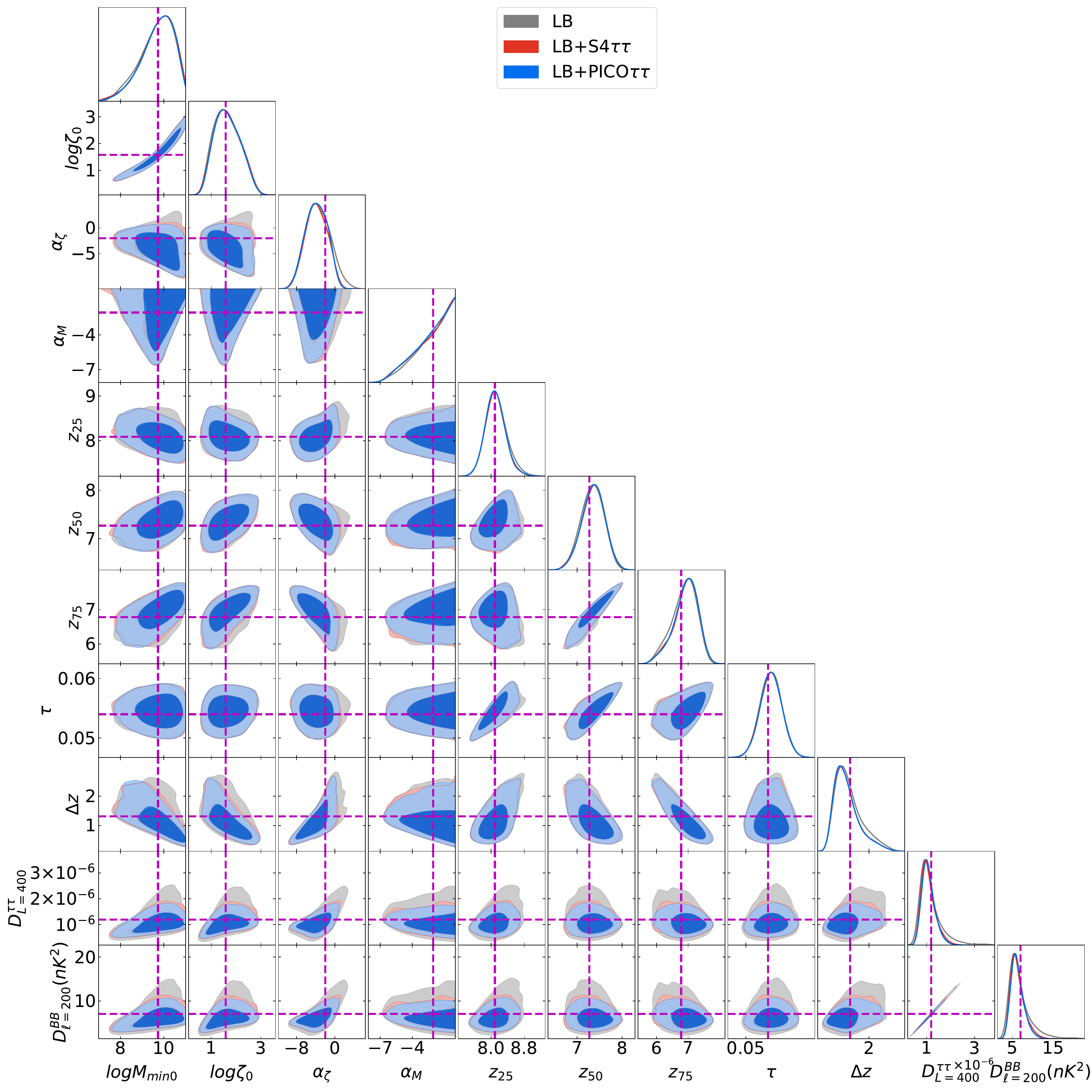}
    \caption{The posterior distribution of free and derived parameters of the reionization model for different combinations of data sets as mentioned in the figure
legend has been presented. The posteriors show both $68\%$ and $95\%$ contours in the two-dimensional posterior plots. The dashed magenta lines denote the input
values used for forecasting.}
    \label{fig:compare4cases-bestfit}
\end{figure*}

\begin{itemize}
    \item \textbf{LB+S4$\tau\tau$:} Projected measurement of $\tau$ with LiteBIRD and $\tau$-power spectrum with CMB-S4 [expected availability $\sim 2030$]
    \item \textbf{LB+PICO$\tau\tau$: }Projected measurement of $\tau$ with LiteBIRD and $\tau$-power spectrum with PICO [expected availability sometime in the next decade]
\end{itemize}
We employ the MCMC sampler in the \texttt{Cobaya} framework \citep{Torrado_2021} to sample the free parameters of our reionization model $\mathbf{\theta}\equiv\rr{\log (\zeta_0),\log M_{\mathrm{min},0},\alpha_\zeta,\alpha_M}$. In Table \ref{tab:paramprior}, the priors for these free parameters, as used during Bayesian inference, are presented. Each set of sample $\mathbf{\theta}$ yields the derived parameters $\rr{\tau,C^{\tau\tau}_{L}}$. We compare these derived parameters to projected mock data sets for each of the above cases and obtain the posteriors on the reionization parameters. The likelihood used in the above analysis has the following form:
    \begin{equation}
    -2\log \mathcal{L}  =  \cc{\frac{\tau-\tau^{\mathrm{obs}}}{\sigma^{obs}_\tau}}^2 +  \sum_{L}\cc{\frac{\bar{C}^{\tau \tau}_{L}-C^{\tau \tau}_{L}}{\Sigma^{\tau\tau}_{L}}}^2 
    \end{equation}
Furthermore, we compare these forecasts with those derived using just the projected measurement of $\tau$ with LiteBIRD in the case denoted by the dataset \textbf{LB} [expected availability $\sim 2030$].  In addition to forecasts on the free parameters of the model, we forecast constraints on parameters associated with reionization history i.e. $z_{25},z_{50},z_{75}$ which corresponds to reionization redshift corresponding to mass-averaged ionization fraction of $Q_{\rm HII}=[0.25,0.50,0.75]$ and $\Delta z=z_{\rm 25}-z_{\rm 75}$, the width of reionization. We also predict constraints on CMB probes of reionization namely $\tau$, patchy $B$-mode power at $\ell=200$, and $\tau$-power spectrum at $L=400$.

For each prescribed case, we present a comparison of the parameter constraints and two-dimensional posterior distribution in Table \ref{tab:fullparamconst_bestfit_move} and Figure \ref{fig:compare4cases-bestfit} respectively. The LB case represents the best possible forecasts on our reionization models with future measurements of $\tau$. As $\tau$ is a measure of the evolution of averaged ionization fraction, $Q_{\rm HII}$, this yields tight constraints on the parameters associated with reionization history. Notably, the error on $z_{50}$ is $ \sim 0.26$ and that on $\Delta z $ is $\sim 0.44$. Further, the ability to provide constraints on the source properties enables us to constrain the patchy picture of reionization. The forecasted error bars on $D^{BB}_{\ell=200}\; ({\rm nK^2})$ and  $D^{\tau\tau}_{L=400}\times 10^6$ are about $\sim 1.89$ and  $\sim 0.83$ respectively.

Comparing forecasts for  LB with LB+S4$\tau\tau$ and LB+PICO$\tau\tau$ data sets we find that the inclusion of $\tau$-power spectrum leads to tighter constraints on  $\alpha_\zeta$, the parameter characterizing how fast ionizing efficiency of the sources evolves with redshift. The correlation between $\alpha_\zeta$ and patchy CMB probes (Figure \ref{fig:compare4cases-bestfit}) implies tighter $\alpha_\zeta$ bounds yield stricter $D^{BB}_{\ell=200}\; ({\rm nK^2})$ constraints, with errors improving from $1.89$ (LB) to $1.55$ (LB+S4$\tau\tau$) and $1.46$ (LB+PICO$\tau\tau$). Similarly, $D^{\tau\tau}_{L=400}\times 10^6$ errors decrease from $0.83$ (LB) to $0.27$ (LB+S4$\tau\tau$) and $0.26$ (LB+PICO$\tau\tau$). Further, a slight improvement in constraints for $\Delta_z$ from error bars of $\sim 0.44$ in LB to $\sim 0.39$ LB+S4$\tau\tau$ and $\sim 0.36$ for LB+PICO$\tau\tau$  is observed, indicating that error bars on reionization history parameters are largely driven by error bars on $\tau$. Still, with improved polarization data set from PICO, we will achieve the tightest constraint on the above parameters. Finally, from the two-dimensional posteriors, a tight correlation is observed between the $\tau$ power spectrum and the patchy $B$-mode signal indicating that both signals arise from the fluctuations in the same free electron density field, constraining the shape of the power spectrum strongly (as indicated with constraints on $D^{\tau\tau}_{L=400}$ and $D^{BB}_{\ell=200}$).

Thus, based on the correlations observed, the 
inclusion of $\tau$-power spectrum signal can be exploited to tighten constraints on patchy $B$-mode polarization, and the scaling relation in Equation \eqref{eq:tautauBB} provides a complementary way to constrain the amplitude of the reionization component.

\section{Joint detection of patchy $B$-mode and primordial $B$-mode by upcoming CMB experiments} \label{sec:atau}
\begin{figure}
	\includegraphics[width=\columnwidth]{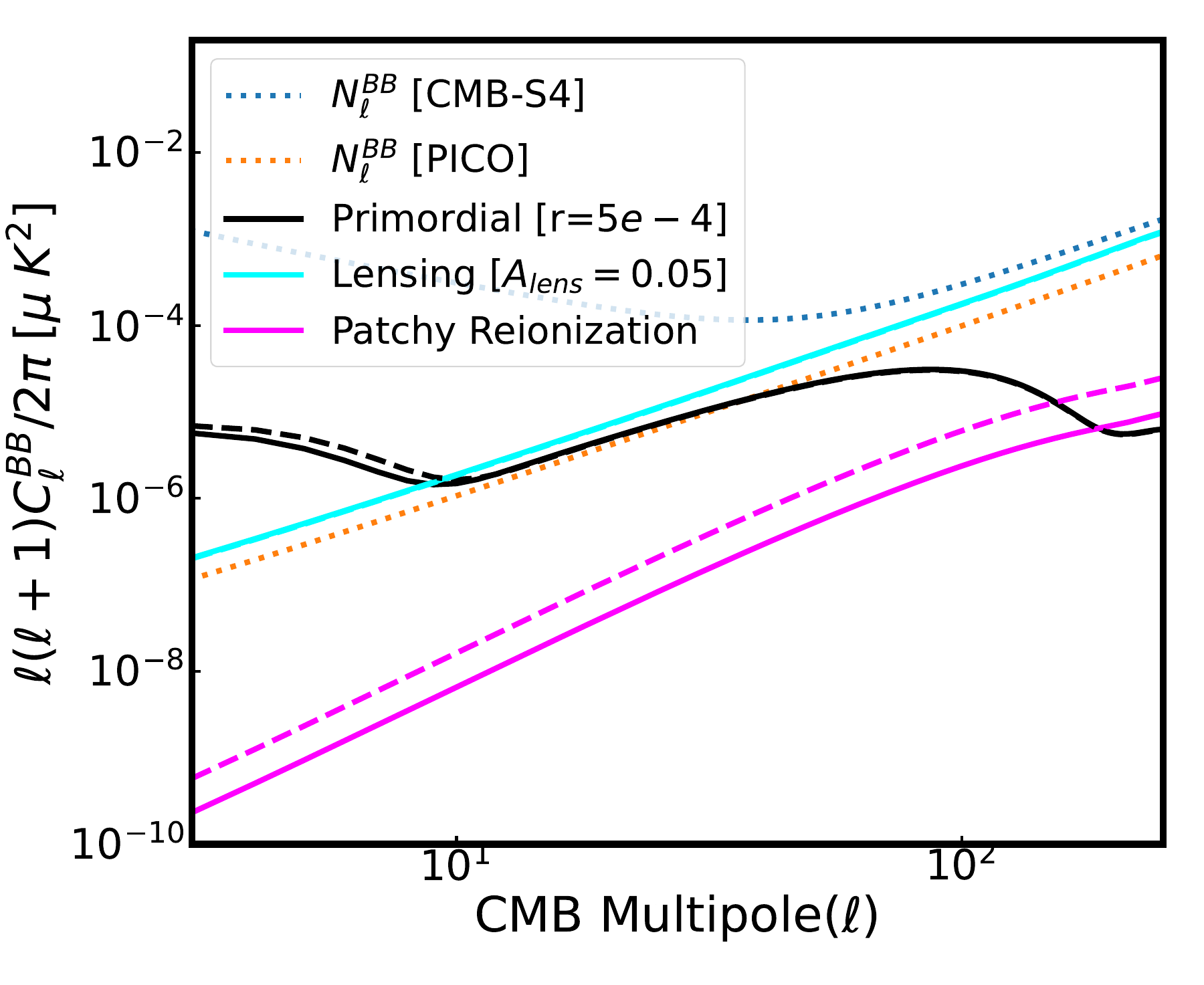}
    \caption{The angular power spectrum of $B$-mode from patchy reionization in
magenta for our fiducial model (solid) and max-BB (dashed) of reionization
has been shown. The corresponding primordial $B$-mode power spectra (black
solid and dashed curves) for the tensor-to-scalar ratio of $5\times10^{-4}$ are also shown.
The cyan curve denotes power due to weak lensing (with lensing amplitude
of $A_{\rm lens} = 0.05$). The dotted lines represent the instrumental noise power
spectra for CMB-S4 ($\ell_{\rm knee}=50;\gamma=-3$) and PICO, in blue and orange curves respectively.}
    \label{fig:noisecurves}
\end{figure}
The Stage-4 CMB experiments will make the most sensitive observations of large-scale $B$-mode polarization to constrain the amplitude of the tensor-to-scalar power spectrum ratio $r$ \citep{2018JLTP..193.1048S,2019arXiv190210541H,2019JCAP...02..056A,2019arXiv190704473A}. The detection of $r$ will be a landmark in establishing the inflationary paradigm \citep[][and references therein]{2016ARA&A..54..227K,2016arXiv160501615C}, the amplitude of which will set the energy scale of inflation. Any constraint on $r$ will also help constraint on the allowed models of inflation \citep{2017JCAP...04..006S,2020A&A...641A..10P,2020EPJC...80.1163Q}. The biggest challenge towards this detection is the $B$-mode contamination coming from the low redshift Universe. The chief sources of this contamination are the synchrotron and dust polarization $B$-mode contribution coming from our galaxy \citep{1997NewA....2..323H,2016ARA&A..54..227K,2020A&A...641A...4P,2020A&A...641A..11P}, the  $B$-mode component arising from weak gravitational lensing by the large-scale structure \citep{2006PhR...429....1L,2014PTEP.2014fB109I}, and the patchy $B$-mode arising from Thomson interaction of CMB photons in the reionization era electron density field \citep{2000ApJ...529...12H}. Efforts are underway to correct for galactic and lensing foreground through multi-wavelength observations of the galactic component \citep{2018A&A...618A.166K} and delensing via internal and external delensing through CMB observations and observations of large-scale structure respectively \citep{2017PhRvD..95j3514N,2017JCAP...05..035C,2017PhRvD..96l3511Y,Millea_2019} .\citep{2017JCAP...05..035C,2022MNRAS.514.5786B}.  \cite{2019MNRAS.486.2042M,2021JCAP...01..003R,2023MNRAS.522.2901J} have studied and concluded the existence of fractional bias arising in the detection of $r$ arising from the mis-modelling of the $B$-mode spectra by neglecting the patchy $B$-mode contribution. In Figure \ref{fig:noisecurves}, we highlight the different contributions to the total $B$-mode spectrum, i.e. the primordial ($r=5e-4,n_t=0$), lensing ($A_{lens}=0.05$), and patchy-$B$-mode for both fiducial and max-BB reionization history. The untangling or ``detau"-ing  of patchy $B$-mode polarization from the primordial has been challenging because of the lack of understanding of the exact reionization details. Consistent with the remarks in \cite{2019MNRAS.486.2042M}, from Figure \ref{fig:noisecurves} we note that within $\ell\sim 2-250$ range, the shape of patchy $B$-mode remains the same only to vary in amplitude depending on the choice of reionization scenario. In light of results presented in the previous section, semi-numerical simulations have allowed us to constraint patchy properties of physical models of reionization based on just CMB observations \citep[and this work]{2019MNRAS.486.2042M,2021MNRAS.501L...7C,2023MNRAS.522.2901J}. This opens up the possibility of inferring both primordial and patchy $B$-mode jointly from the CMB data.

In this Section, we explore the use of estimators ($B$-mode power spectrum and $\tau$ power spectrum) of patchy reionization to recover the reionization $B$-mode component, allowed by the sensitivities of Stage-4 CMB experiments. If we can constrain the amplitude of the patchy $B$-mode signal, it will serve as a zeroth-order extraction of the patchy $B$-mode signal from the total $B$-mode power spectrum in an effort towards unbiased detection of $r$. In this effort, we attempt to simultaneously constrain the amplitude of the reionization contribution, $A_\tau$, and the tensor-to-scalar power spectrum ratio, $r$.\\

\subsection{Formalism to extract the amplitude of patchy-$B$ mode spectrum}
For this analysis, we consider, the mock $B$-mode spectrum given by 
\begin{equation}\label{eq:Ataumock}
    \tilde C^{BB}_\ell=\tCBB{prim}+A_{lens}\tCBB{lens}+\tCBB{reion}
\end{equation}
where the primordial and lensing $B$-mode signal, $\tCBB{prim},A_{lens}\tCBB{lens}$ is calculated using the Modified CAMB in our framework, discussed earlier in Section \ref{sec:signalandsim}. This allows for the evaluation of CMB anisotropies in a self-consistent fashion with the reionization history predicted by our model. $\tCBB{reion}$ denotes the patchy $B$-mode spectrum corresponding to the choice of the model of reionization. In this study, we consider the fiducial and the max-$BB$ model of reionization to construct the mock $B$-mode data set. The primordial $B$-mode mock is constructed using $r=5\times 10^{-4}$ and $n_t=0$.

We define the model spectrum such that the contribution of patchy $B$-mode, $C^{BB, {\rm reion}}_{\ell}$, in the total model $B$-mode spectrum, $C^{BB}_{\ell}$, is determined by the term $A_\tau$ and indicates the amplitude of patchy $B$-mode corresponding to the fiducial model of reionization.
\begin{equation}
    C^{BB}_\ell=\CBB{prim}+A_{lens}\CBB{lens}+A_\tau C^{BB, {\rm reion}}_{\ell, {\rm fid}}
\end{equation}
In the model spectrum, we evaluate the primordial and lensing $B$-mode signal, $\CBB{prim}$ and $A_{lens}\CBB{lens}$ through the default CAMB routine.

Further, the $\tau$-power spectrum signal at angular scales (smaller than the reionization bump) is related to the scattering $B$-mode spectrum
in Equation \eqref{eq:tautauBB}. Therefore, when using the projected $\tau$-power spectrum data set, our model $\tau$-power spectrum is approximated as 
\begin{equation}
   C^{\tau\tau}_{L}\approx\frac{100}{3}\frac{1}{Q^{2}_{\rm rms}}A_\tau C^{BB, {\rm reion}}_{L, {\rm fid}}e^{2\tau}
\end{equation}
We present forecasts for jointly recovering $A_\tau$ and $r$ with CMB-S4 and PICO considering two combinations of projected $\tau$, $\tau$-power spectrum and $B$-mode signal data sets:
\begin{itemize}
    \item Case \textbf{LB +S4$BB$ + S4$\tau\tau$}: LiteBIRD ($\tau$)  + CMB-S4 ($B$-mode signal) + CMB-S4 ($\tau$-power spectrum) 
    \item Case \textbf{LB + PICO$BB$ + PICO$\tau\tau$}: LiteBIRD ($\tau$) + PICO ($B$-mode signal) + PICO ($\tau$-power spectrum) 
\end{itemize}
\begin{table}
    \def\arraystretch{1.6}
    \centering
    \caption{Specifications of priors for the free parameters used in the joint recovery of the amplitude of the patchy $B$-mode signal and tensor-to-scalar power spectrum ratio, $r$.}
    \begin{tabular}{|c|c|c|}
    \hline
    Parameter & range & nature \\
    \hline
    $r$     & (0,$\infty$) & uniform\\
    $\tau$  & (0,$\infty$) & uniform \\
    $A_\tau$ & (0,$\infty$) & uniform\\
    \hline
    \end{tabular}
    \label{tab:Atauprior}
\end{table}
\begin{table*}
    \centering
    \def\arraystretch{1.6}
    \caption{Parameter forecasts ($68\%$ limits) on free parameters of the model from the MCMC analysis of recovering  $r$ and  $A_\tau$, for the fiducial and max-$BB$ case corresponding to the combination of CMB data sets. The second column shows the input value used to construct the mock data based on which forecasts are made. Constraints on parameter $r$ is presented as  $\cc{r^{\sigma_+}_{\sigma_-}}\times 10^3$. We present an estimated measure of the significance of the recovery of $A_\tau$ for each case, represented by the parameter $\mathbf{A_\tau/\sigma_{A_\tau}}$.}
    \begin{tabular}{|c|c|c|c|c|c|c|}
    \hline
    && \multicolumn{2}{c}{LiteBIRD$\tau$ + CMB-S4 data sets} & &\multicolumn{2}{c}{LiteBIRD$\tau$ + PICO data sets}\\
    \cline{3-4} \cline{6-7}
    Parameters  & Input   &  LB+S4$BB$&  LB+S4$BB$+S4$\tau\tau$&  & LB+PICO$BB$& LB+PICO$BB$+PICO$\tau\tau$\\
    \hline
    {Delensing at 95\%} && \multicolumn{5}{c}{\textit{fiducial Reionization model} }\\
    $r \times 10^3$  & $0.5$ &$0.456^{+0.107}_{-0.106}$    &$0.502^{+0.096}_{-0.102}$ && \pc{0.493}{0.037}{0.037}   & $0.498^{+0.034}_{-0.034}$ \\
    $\tau$ & $0.0540$    & $0.0542^{+0.0020}_{-0.0021}$  & $0.0540^{+0.0020}_{-0.0020}$  & & \pc{0.0541}{0.0019}{0.0019} (0.0539) & $0.0539^{+0.0019}_{-0.0018}$ \\
    $A_\tau$ & $1.0$ &  $<3.11$  &  $0.99^{+0.43}_{-0.45}$ && \pc{1.36}{0.50}{1.19}   & $1.01^{+0.32}_{-0.32}$ \\
    S/N = \mathbf{A_\tau/\sigma_{A_{\tau}}}&&-&\textbf{2.25}&&\textbf{1.60}&\textbf{3.16}\\
    \hline
     {Delensing at 85\%} &&\multicolumn{5}{c}{\textit{max-BB Reionization model}}\\
    $r \times 10^3$  & $0.5$ & \pc{0.450}{0.178}{0.179} ) & \pc{0.510}{0.159}{0.160}  &&\pc{0.485}{0.074}{0.074}  & \pc{0.494}{0.067}{0.068}    \\
    $\tau$ & $0.0627$     &  \pc{0.0630}{0.0019}{0.0019}  & \pc{0.0626}{0.0019}{0.0018}  && \pc{0.0629}{0.0020}{0.0019}   & \pc{0.0626}{0.0018}{0.0017}  \\
    $A_\tau$ & $2.58$    & $<5.39$  & \pc{2.56}{0.64}{0.64} & &\pc{3.38}{1.33}{2.83}  & \pc{2.61}{0.53}{0.54}  \\
    S/N = \mathbf{A_\tau/\sigma_{A_{\tau}}}&&-&\textbf{4.00}&&\textbf{1.63}&\textbf{4.83}\\   
     {Delensing at 95\%} && \\
    $r \times 10^3$  & $0.5$ & \pc{0.487}{0.107}{0.109}  & \pc{0.505}{0.102}{0.102}  &&\pc{0.495}{0.040}{0.039}   & \pc{0.497}{0.034}{0.036}    \\
    $\tau$ & $0.0627$     & \pc{0.0628}{0.0019}{0.0019}  & \pc{0.0627}{0.0018}{0.0019}  && \pc{0.0627}{0.0020}{0.0019}   & \pc{0.0626}{0.0018}{0.0018}   \\
    $A_\tau$ & $2.58$     & \pc{3.08}{1.38}{2.31}  & \pc{2.57}{0.44}{0.45} & &\pc{2.70}{1.02}{1.13}  & \pc{2.60}{0.34}{0.33}  \\
    S/N = \mathbf{A_\tau/\sigma_{A_{\tau}}}&&\textbf{1.85}&\textbf{5.71}&&\textbf{2.50}&\textbf{7.65}\\
    \hline
    \end{tabular}
    \label{tab:bf-alldatasets}
\end{table*}

In the analysis, we use 3 parameters: r, $\tau$, and $A_\tau$ to entirely describe our model spectrum, hence, these are the free parameters. The priors for the free parameters are provided in Table \ref{tab:Atauprior}. The Likelihood function used in this analysis is given as:
\begin{equation}
    -2\log \mathcal{L}  =  \cc{\frac{\tau-\tau^{\mathrm{obs}}}{\sigma^{obs}_\tau}}^2 +  \sum_{\ell}\cc{\frac{\bar{C}^{BB}_{\ell}-C^{BB}_{\ell}}{\Sigma^{BB}_{\ell}}}^2+  \sum_{L}\cc{\frac{\bar{C}^{\tau\tau}_{L}-C^{\tau\tau}_{L}}{\Sigma^{\tau\tau}_{L}}}^2 
\end{equation}
To emphasize the contribution of the $\tau$-power spectrum dataset in the recovery of $r$ and $A_\tau$, we present additional forecasts using just the projected $\tau$ and projected $B$-mode signal. We refer to these scenarios as Case \textbf{ LB +S4$BB$} and Case \textbf{ LB + PICO$BB$}. 

\begin{figure}

	\includegraphics[width=\columnwidth]{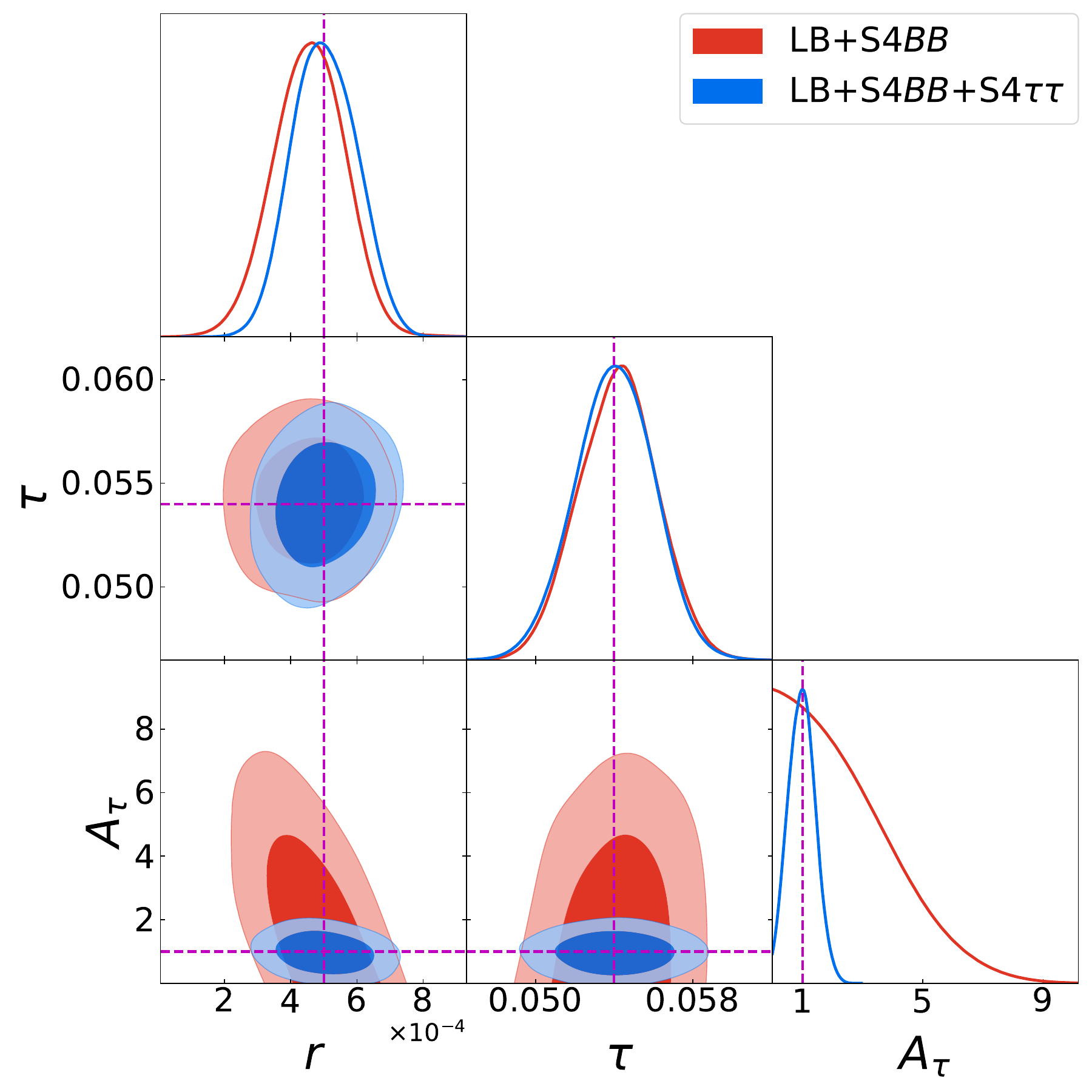}
    \caption{Posterior distribution of free parameters from the MCMC analysis of recovering $r$ and $A_\tau$ for the fiducial case  corresponding to the combination of CMB-S4's $\tau$-power spectrum and $B$-mode data sets with LiteBIRD's projected measurement of $\tau$ (refer figure legend). The posteriors show both $68\%$ and $95\%$ contours in the two-dimensional posterior panels. The dashed magenta lines denote the input
values used for forecasting.}
    \label{fig:bf-S4data}
\end{figure}

\begin{figure}

	\includegraphics[width=\columnwidth]{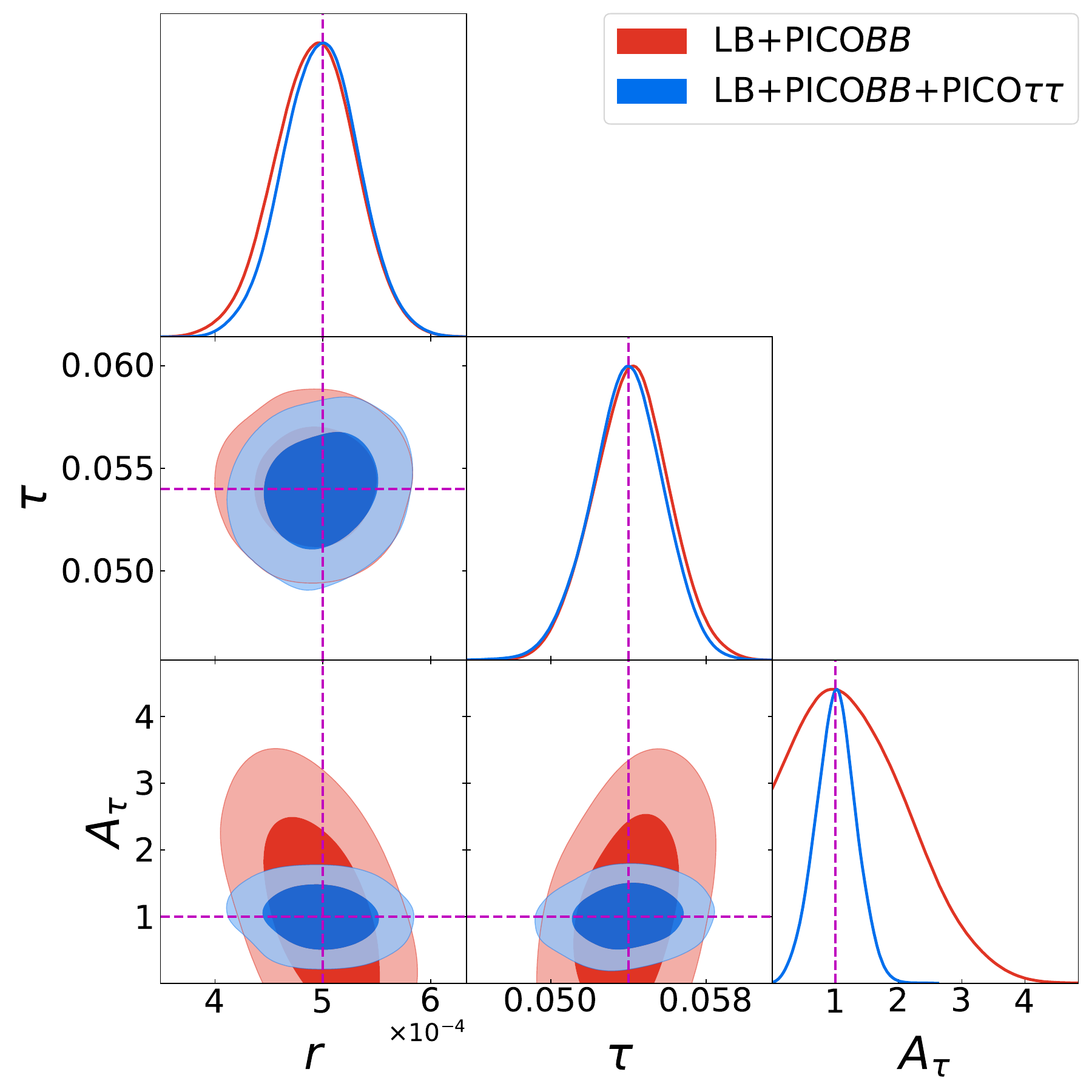}
    \caption{Posterior distribution of free parameters from the MCMC analysis of recovering $r$ and  $A_\tau$ for the fiducial case  corresponding to the combination of PICO's $\tau$-power spectrum and $B$-mode data sets with LiteBIRD's projected measurement of $\tau$ (refer figure legend). The posteriors show both $68\%$ and $95\%$ contours in the two-dimensional posterior panels. The dashed magenta lines denote the input
values used for forecasting.}
    \label{fig:bf-picodata}
\end{figure}

\begin{figure}

	\includegraphics[width=\columnwidth]{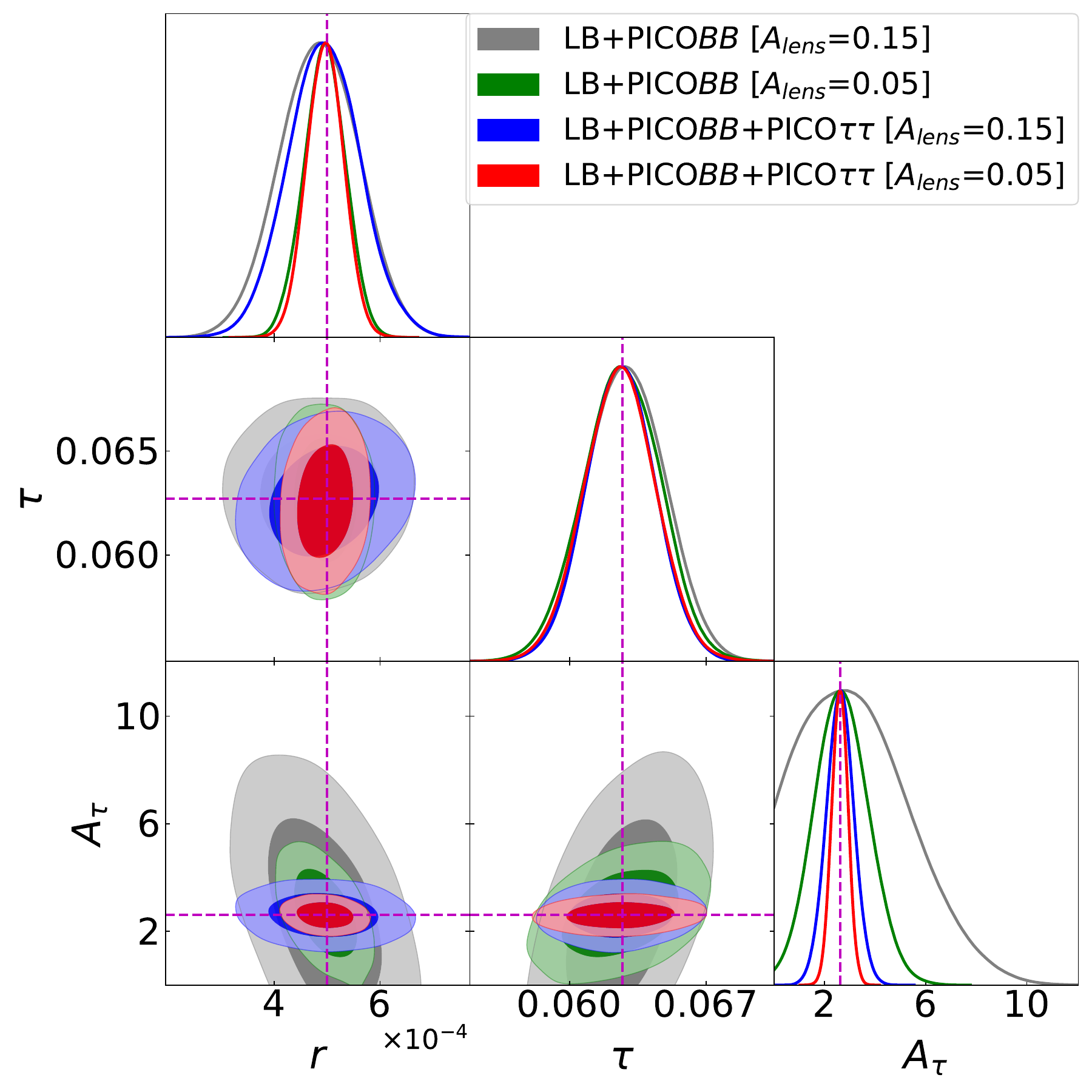}
    \caption{Posterior distribution of free parameters from the MCMC analysis of recovering $r$ and $A_\tau$ for the max-BB case  corresponding to the combination of PICO's $\tau$-power spectrum and $B$-mode data sets with LiteBIRD's projected measurement of $\tau$ (refer figure legend). The posteriors show both $68\%$ and $95\%$ contours in the two-dimensional posterior panels. The dashed magenta lines denote the input
values used for forecasting.}
    \label{fig:bf-3sigmapicodata}
\end{figure}

\subsection{Bayesian forecasts on the amplitude of patchy-$B$ mode spectrum}

Considering the case of the fiducial model of reionization, forecasts for our model's free parameter corresponding to the combination of CMB data sets are presented in Table \ref{tab:bf-alldatasets}. The two-dimensional posterior distributions are illustrated in Figure \ref{fig:bf-S4data} (for LB+CMB-S4) and Figure \ref{fig:bf-picodata} (for LB+PICO). We find that with the combination of the $B$-mode signal with LiteBIRD's projected $\tau$ for the scenarios LB+S4$BB$ and LB+PICO$BB$, PICO can facilitate a $68\%$ constraint on $A_\tau$ at \pc{1.356}{0.503}{1.189}, roughly detecting it at $1.6\sigma$, while placing an upper limit of 3.11 at the $68\%$ confidence level with the use of CMB-S4 data sets. 
The nature of correlations (illustrated by red contours in Figure \ref{fig:bf-S4data} and Figure \ref{fig:bf-picodata}) observed is in line with expectations. A slight negative correlation is observed between the amplitude of the patchy-$B$ mode signal and the tensor-to-scalar power spectrum $r$, indicating that in models where $A_\tau$ is neglected, a higher $r$ is inferred. The slight positive correlation between $A_\tau$ and $\tau$ is indicative of higher preference of $A_\tau$, as power in models with higher $\tau$ is damped by a factor of $e^{-2\tau}$. The incorporation of the $\tau$-power spectrum datasets in the cases of LB+S4$BB$+S4$\tau\tau$ and LB+PICO$BB$+PICO$\tau\tau$ breaks these correlations, evident in blue contours in Figure \ref{fig:bf-S4data} and \ref{fig:bf-picodata} as a consequence of higher-signal to noise in detecting the $\tau$-power spectrum signal. This comparatively improves our capability to estimate the patchiness during the reionization era, providing tighter constraints on $A_\tau$. This is evident in Table \ref{tab:bf-alldatasets}, where the inclusion of $\tau$-power spectrum data set with CMB-S4 and PICO can impose tighter constraints on $A_\tau$. $A_\tau$ is detected at roughly $\sim 2.25 \sigma$ and $\sim 3.16 \sigma$ with CMB-S4 and PICO datasets respectively. This makes PICO a prospective experiment to achieve the first $3\sigma$ detection of the patchy $B$-mode signals, a crucial development in the attempt to achieve unbiased measurement of $r$. Constraining $A_\tau$ through $\tau$-power spectrum data set also leads to improvement in detection significance of $r$ by $0.79\sigma$ for LiteBIRD $\tau$+CMB-S4 data sets and $1.32\sigma$ for LiteBIRD$\tau$+PICO data sets.

Considering the case of max-$BB$ models of reionization,  which represents recovering $A_\tau$ for an extreme case of reionization, forecasts corresponding to the combination of data sets and delensing scenarios are presented in Table \ref{tab:bf-alldatasets}. We observe a similar trend of improvement, as discussed in the case of the fiducial model, in constraining $A_\tau$ with the inclusion of $\tau$-power spectrum data set. Even with a delensing scenario of 85\%, we will be able to achieve notable detection with CMB-S4, making a $\sim 4\sigma$ detection, while PICO would make a detection of $\geq 4.8 \sigma$. The detectability further improves with improved delensing efficiencies, going as high as $7\sigma$ for PICO observations at $95\% $ delensing. It is important to note that delensing is the process of reducing the lensing-induced $B$-modes to expose other $B$-mode components. Therefore, better delensing significantly improves our ability to detect $A_\tau$. As the nature correlations for max-BB cases are similar to the correlations observed for the case of fiducial model of reionization in Figure \ref{fig:bf-picodata} and Figure \ref{fig:bf-S4data}, we only show the two-dimensional posteriors of the free parameters corresponding to the case of PICO data sets in Figure \ref{fig:bf-3sigmapicodata} which correspond to the highest detection significance of $A_\tau$. The nature of correlations, i.e. positive correlation between $A_\tau$ and $\tau$ and negative correlation between $A_\tau$ and $r$ is observed for the combination of projected $\tau$ and $B$-mode data sets, consistent with our previous inference from Figure \ref{fig:bf-S4data} and Figure \ref{fig:bf-picodata}.

The above cases illustrate the first joint constraint on $r$ and $A_\tau$ within the realm of physically motivated models of reionization. Stage-4 CMB experiments, through their $B$-mode observations, will open a window to detect both the primordial and patchy $B$-mode components. We find that $B$-mode estimators in conjunction with estimators for off-diagonal correlations will enable the first detection $3\sigma$ of patchy $B$-mode power spectrum through polarization observations with PICO. This advancement is pivotal for uncovering patchy properties of reionization and represents a step towards achieving unbiased constraints on $r$. Apart from providing a measure of patchiness, the ability to extract $B$-modes will allow an improved measure of $r$, evident through its increased detection significance.  The degradation of the S/N in the presence of foreground contamination is likely to be negligible if the spectral dependence of the foregrounds can successfully mitigate the contamination using multiple frequency channels.

\section{Discussion AND Conclusion}
The patchiness in the process of reionization introduces secondary imprints in the CMB. One of the primary consequences is the large-scale, patchy $B$-mode signature in the CMB polarization. Neglecting this signature in modelling the total $B$-mode signal will introduce a fractional bias in the inference of the tensor-to-scalar power spectrum ratio, $r$. Mitigating this foreground via templates based on multi-frequency observation is futile, as the patchy $B$-mode has the same frequency dependence as the actual CMB. The situation is further plagued by uncertainties in reionization modelling. An epoch of reionization that includes larger degree-scale fluctuations leads to a more pronounced large-scale $B$-mode signal, complicating the possibility of achieving an unbiased measurement of $r$.

Data-driven models of reionization and estimators of patchiness in the electron density field, facilitating accurate predictions of the strength of patchy $B$-mode signal, are therefore crucial for an effective resolution of this challenge. In this study, we employ a physically motivated model of reionization which enables the detailed tracking of the electron density's patchiness at various redshifts. We investigate the potential for jointly  constraining the reionization $B$-mode and the primordial $B$-mode signal via a conjunction of $B$-mode and $EB$ estimator, in light of unprecedented sensitivities to be achieved with the Stage 4 CMB experiments like CMB-S4 and PICO. 
\begin{figure}
	\includegraphics[width=\columnwidth]{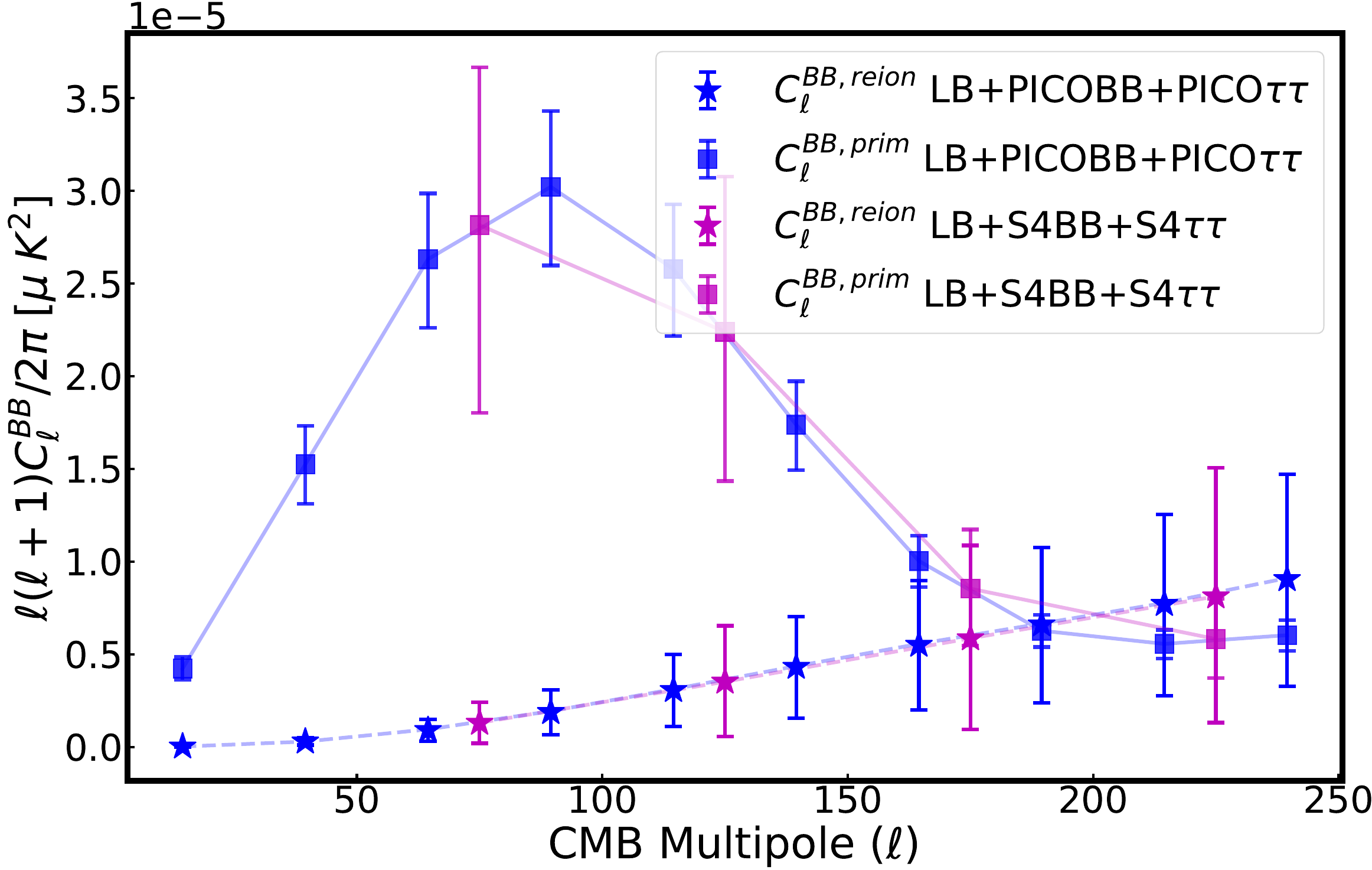}
    \caption{The angular power spectrum of $B$-mode from patchy reionization ($D^{BB,reion}_{\ell}$) and primordial gravitational waves ($D^{BB,prim}_{\ell}$) with $2\sigma$ limits have been presented corresponding to forecasts of  $A_\tau$ (refer Table \ref{tab:bf-alldatasets}) for the cases LB+S4$BB$+S4$\tau\tau$ and LB+PICO$BB$+PICO$\tau\tau$ with fiducial model of reionization.}
    \label{fig:2sigmalimits}
\end{figure}
We find that for our four-parameter model of reionization, the inclusion of projected $\tau$-power spectrum detected with CMB-S4 and PICO will lead to tighter estimates of the patchy $B$-mode signal compared to constraints forecasted with just the sky-averaged $\tau$ measurements with LiteBIRD. This is essentially a consequence of the $\tau$-power spectrum and the $B$-mode signal being sourced through the same fluctuations in the electron density field, as  evidenced through the strong correlation observed in the posteriors for $\tau$ and $B$-mode power spectrum. We exploit this complementary way of constraining the patchy $B$-mode signal to detect the strength of the signal by CMB-S4 and PICO, studying it in regard to a fiducial model and a max-BB model of reionization. While the fiducial model corresponds to the best-fit model achieved when confronting our four-parameter model of reionization with the current observables of CMB, the max-BB corresponds to a model which produces the maximum $B$-mode signal allowed by current CMB estimates. Sensitive observations through PICO would allow the first $3\sigma$ detection of the reionization $B$-mode signal. This increases to a maximum of $7\sigma$ detection of the signal if the true model of reionization were the max-BB model. Improved detection of $A_\tau$, the amplitude of patchy $B$-mode, is not only vital for understanding the patchy reionization process but also leads to enhanced detection significance of the tensor-to-scalar power spectrum ratio $r$. This enhancement is critical for Stage-4 CMB experiments, which are designed to make high-fidelity detection of $r$ and constrain the diverse range of inflationary scenarios. We summarize our findings through Figure \ref{fig:2sigmalimits}, where we present the projected $2\sigma$ error bars for the detection of primordial and patchy $B$-mode components in the context of upcoming observations with  Stage-4 CMB experiments.

This work marks the first study of the detectability of the patchy $\tau$-power spectrum signal, based on realistic and physically motivated numerical models of reionization. In comparison to previous works \citep{Dvorkin:2008tf,2018JCAP...05..014R} which used simplistic spherical bubble-based prescription, we employ a  semi-numerical model of reionization to simulate the patchy ionization fields. In addition to studying the role of $\tau$-power spectrum in improving our knowledge of the patchy reionization process, we illustrate its role in the extraction of the patchy $B$-mode signal from CMB $B$-mode observations. This is significant, as it paves the way towards "detau"-ing the primordial $B$-mode signal from the patchy reionization component, ensuring an unbiased measurement of the tensor-to-scalar power spectrum ratio $r$ free from at least one extra-galactic foreground contamination having the same frequency spectrum as CMB.


\section*{Acknowledgments}
DJ and TRC acknowledge support of the Department of Atomic
Energy, Government of India, under project no. 12-R\&D-TFR-
5.02-070. The work of SM is a part of the $\langle \texttt{data|theory}\rangle$ \texttt{Universe-Lab} which is supported by the TIFR and the Department of Atomic Energy, Government of India. 

 \section*{Data Availability}
A basic version of the semi-numerical code SCRIPT for generating the
ionization maps used in the paper is publicly available at \url{https://
bitbucket.org/rctirthankar/script}. Any other data related
to the paper will be shared on reasonable request to the corresponding
author (DJ).
 
\bibliographystyle{mnras}
\bibliography{example}

\appendix

\section{Forecasts on Free Parameters in the Absence of Primordial Gravitational Waves}

\begin{table}
    \centering
    \def\arraystretch{1.6}
    \caption{Parameter forecasts ($68\%$ limits) on free parameters of the model from the MCMC analysis of recovering $A_\tau$ for the fiducial case corresponding to the combination of CMB data sets and considering mock $r=0$. The second column shows the input value used to construct the mock data based on which forecasts are made. Constraints on parameter $r$ presented as  $\cc{r^{\sigma_+}_{\sigma_-}}\times 10^3$.}
    \begin{tabular}{|c|c|c|c|}
    \hline
    Parameters  & Input    & LB+S4$BB$+S4$\tau\tau$& LB+PICO$BB$+PICO$\tau\tau$\\
    \hline
    $r\times 10^3$ & 0 & $<0.086$  & $<0.027$ \\
    $\tau$  & 0.054  & $0.0540^{+0.0020}_{-0.0019}$  & $0.0542^{+0.0019}_{-0.0018}$ \\
    $A_\tau$ &1.00  &  $0.960^{+0.416}_{-0.440}$ & $0.963^{+0.308}_{-0.308}$ \\
    S/N = \mathbf{A_\tau/\sigma_{A_\tau}} & & \mathbf{2.24} & \mathbf{3.13}\\
    \hline
    \end{tabular}
    \label{tab:Ataubf:r=0}
\end{table}

\begin{figure}

	\includegraphics[width=\columnwidth]{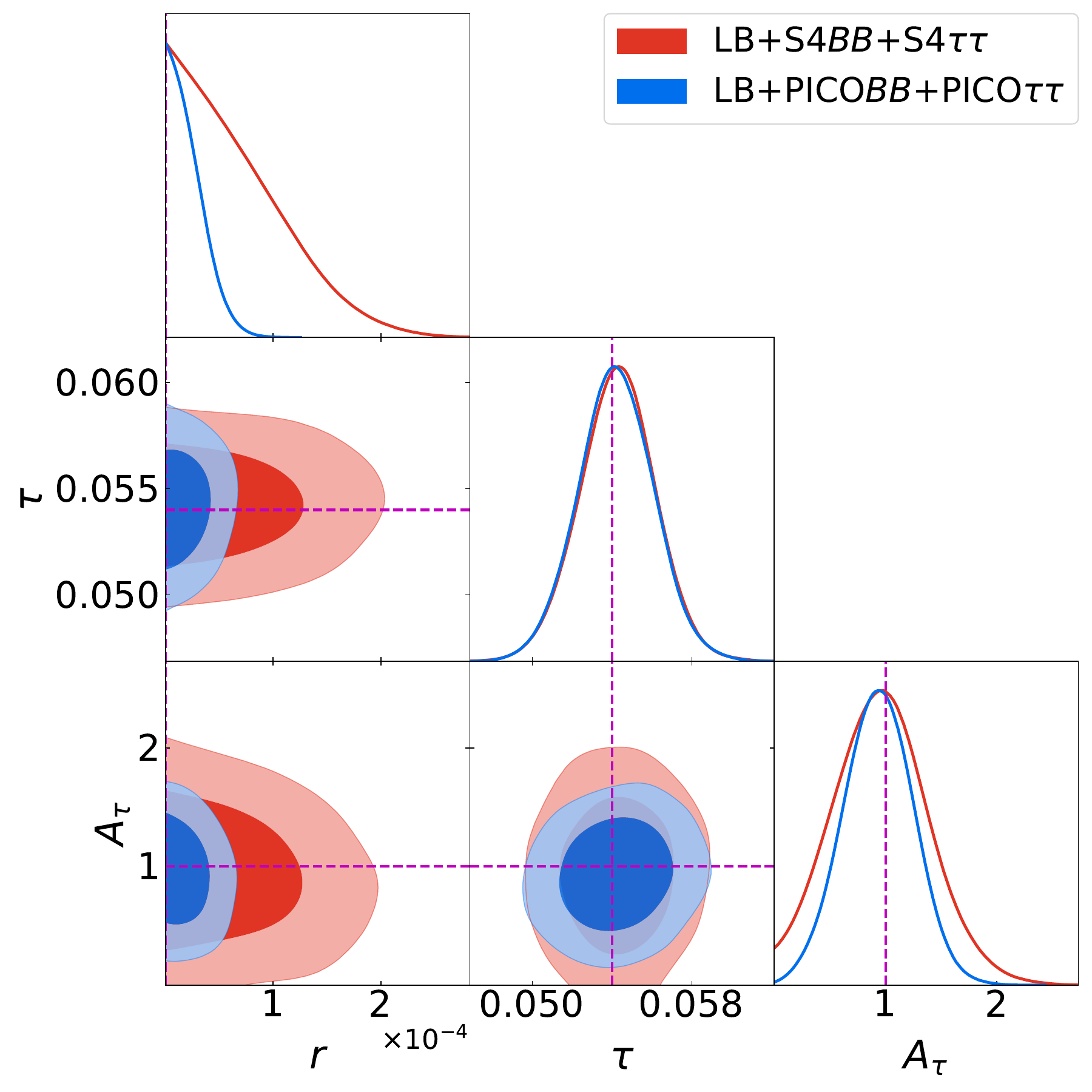}
    \caption{Posterior distribution of free parameters from the MCMC analysis of recovering $A_\tau$ for the fiducial case corresponding to the combination of CMB data sets (refer figure legend) and considering mock $r=0$. The posteriors show both $68\%$ and $95\%$ contours in the two-dimensional posterior plots. The dashed magenta lines denote the input
values used for forecasting.}
    \label{fig:bf-r0}
\end{figure}

Here we present the case, considering $r=0$ in the mock power spectrum i.e. extraction of $A_\tau$ in the absence of primordial gravitational waves. We present forecasts for the case LB+S4$BB$+S4$\tau\tau$ and LB+PICO$BB$+PICO$\tau\tau$ considering the fiducial model of reionization and $95\%$ delensing. The comparison for parameter constraints from the MCMC analysis has been presented in Table \ref{tab:Ataubf:r=0} and the two-dimensional posterior distribution has been presented in Figure \ref{fig:bf-r0}. Intriguingly, the results reveal that the detection of the patchy $B$-mode with improved error bars, compared to the forecasts for $r=5\times 10^{-4}$ presented in Table \ref{tab:bf-alldatasets}, is feasible with CMB-S4 and PICO in the absence of any primordial gravitational wave signal.

\bsp	
\label{lastpage}
\end{document}